\documentclass[aps,prb,twocolumn,showpacs,citeautoscript,longbibliography]{revtex4-1}
\usepackage{graphicx}
\usepackage{bm}
\usepackage{amsmath}
\usepackage{amssymb}
\usepackage{wasysym}
\usepackage{amsfonts}
\usepackage[colorlinks=true, linkcolor=blue, citecolor=blue, urlcolor=blue, linktoc=page, bookmarks=false, pdfstartview={FitH}, pdfborder={0 0 0.0 [3 3]}]{hyperref} 
\usepackage{mathtools}
\usepackage[usenames,dvipsnames]{xcolor}
\hypersetup{pdfborder=0 0 0,colorlinks=true, citecolor=blue, linkcolor=blue}

\begin{document}

\title{Calculating the transport properties of magnetic materials from first-principles including thermal and alloy disorder, non-collinearity and spin-orbit coupling}

\author{Anton A. Starikov}
\author{Yi Liu}
\altaffiliation[Present address: ]{The Center for Advanced Quantum Studies and Department of Physics, Beijing Normal University, 100875 Beijing, China}
\author{Zhe Yuan}
\altaffiliation[Present address: ]{The Center for Advanced Quantum Studies and Department of Physics, Beijing Normal University, 100875 Beijing, China}
\email[Corresponding author: ]{zyuan@bnu.edu.cn}
\author{Paul J. Kelly}
\email[Corresponding author: ]{P.J.Kelly@utwente.nl}
\affiliation{Faculty of Science and Technology and MESA$^+$ Institute
for Nanotechnology, University of Twente, P.O. Box 217, 7500 AE
Enschede, The Netherlands}

\date{\today~version rv1d}

\begin{abstract}
A density functional theory based two-terminal scattering formalism that includes spin-orbit coupling and spin non-collinearity is described. An implementation using tight-binding muffin-tin orbitals combined with extensive use of sparse matrix techniques allows a wide variety of inhomogeneous structures to be flexibly modelled with various types of disorder including temperature induced lattice and spin disorder. The methodology is illustrated with calculations of the temperature dependent resistivity and magnetization damping for the important substitutional disordered magnetic alloy Permalloy (Py), Ni$_{80}$Fe$_{20}$. Comparison of calculated results with recent experimental measurements of the damping (including its temperature dependence) indicates that the scattering approach captures the most important contributions to this important property. 
\end{abstract}


\maketitle

\section{Introduction}
\label{sec:intro}
As long as device dimensions were much larger than the spin-flip diffusion length of the constituent materials, the effect of the electron spin on transport properties went largely undetected. When attention focussed on magnetic materials in thin film and multilayer form, new properties such as interface magnetic anisotropy and oscillatory exchange coupling emerged, culminating in the discovery of giant magnetoresistance \cite{Baibich:prl88, Binasch:prb89} (GMR) almost 30 years ago \cite{[See the collections of articles ]UMS,*SC12}. This heralded the emergence of the field of spintronics \cite{Bader:arcmp10}, which exploits the spin of electrons in addition to the charge used in conventional electronics, triggering a flood of new discoveries including tunneling magnetoresistance (TMR) \cite{Moodera:prl95, Yuasa:jpd07}, spin-transfer torque (STT) \cite{Slonczewski:jmmm96, Berger:prb96, Ralph:jmmm08}, the spin Hall effect \cite{Hoffmann:ieeem13, Sinova:rmp15} the spin Seebeck effect, etc.\cite{[See the collections of articles ]SpinCurrent12, *Stamps:jpd14} Spin-dependent electron transport manifests itself on microscopic length scales in magnetically inhomogeneous systems such as magnetic bilayers, multilayers and magnetic textures where interface and finite size effects are dominant. As important as the fundamental physics of spin-dependent transport are the applications that spintronics makes possible. The GMR effect allowed magnetic read heads to be miniaturised and led to an explosion in the density of data that could be stored on a hard disk. The TMR effect in magnetic tunnel junctions (MTJs) forms the basis for new forms of non-volatile storage, magnetic random access memories (MRAM); MTJs are also used as sensor elements in read heads. STT makes it possible to write information in MRAMs more efficiently leading to STT-RAMs\cite{Akerman:sc05, Kryder:ieeem09} or to make microwave frequency STT oscillators (STOs) where the injected spin forces a magnetisation to precess with GHz frequency \cite{Berger:prb96, Ruotolo:natn09, Slavin:natn09}. Passage of a spin-polarised current can also cause a domain wall to move, which is the principle behind a form of shift register called ``racetrack memory'' \cite{Parkin:sc08, Thomas:sc10}.

The search for new and improved kinds of magnetic storage provided another focus of attention in the field of spintronics: magnetisation dynamics in response to external fields and currents in nanoscale systems \cite{Tserkovnyak:rmp05}. The physics of such devices involves two major contributions: (i) spin-dependent scattering of electrons in bulk materials and at interfaces, and (ii) spin-non-conserving scattering of electrons because of spin-orbit coupling (SOC) when spin is no longer a good quantum number, or because of magnetic disorder. A breakdown of spin-conservation is essential for spin-relaxation processes that are described with material dependent time- and length-scales, conventionally the Gilbert damping parameter $\alpha$ and the spin-flip diffusion length $l_{\rm sf}$, respectively. Predicting and controlling these properties is very important for understanding and designing new spintronic devices leading to numerous experimental \cite{Bhagat:prb74, Heinrich:jap79, Mizukami:jmmm01, Mizukami:jmmm02, Lagae:jmmm05, Rantschler:ieeem05, Inaba:ieeem06, Oogane:jjap06} and theoretical \cite{Kambersky:jmmm92, Gilmore:prl07, Gilmore:jap08, Kambersky:prb07} material-dependent studies  on the subject. The development of a new theoretical framework \cite{Brataas:prl08, *Brataas:prb11} for calculating magnetization damping and its implementation in the framework of density functional theory \cite{Starikov:prl10, LiuY:prb11, Ebert:prl11, Mankovsky:prb13, Turek:prb15} has motivated systematic reinvestigation of the damping in alloys \cite{Schoen:natp16, Schoen:prb17b} and of the temperature dependence of damping in permalloy \cite{Zhao:scr16} allowing quantitative confrontation of theory and experiment without invoking adjustable parameters such as the relaxation time in the torque correlation method (TCM)  \cite{Kambersky:cjp76, Gilmore:prl07, Gilmore:jap08, Kambersky:prb07}.

In this paper we describe in detail a method we recently used to calculate the resistivity $\rho$, spin flip diffusion length (SDL), and Gilbert damping parameter for Ni$_{1-x}$Fe$_x$ substitutional alloys \cite{Starikov:prl10}, the resistivity and damping for the itinerant ferromagnets Fe, Co and Ni with thermal disorder \cite{LiuY:prb11}, the resistance \cite{YuanZ:prl12} and anisotropic damping  \cite{YuanZ:prl14} of magnetic domain walls, the nonadiabatic STTs in ballistic systems \cite{YuanZ:prb16}, interface-enhanced damping \cite{LiuY:prl14}, thermal disorder effects in transport \cite{LiuY:prb15} and a  novel interface spin Hall effect \cite{WangL:prl16}. It extends earlier work \cite{Xia:prb01, Xia:prb06, Zwierzycki:pssb08} by  including SOC and non-collinearity.

\begin{figure}[tbp]
  \centering
\includegraphics[width=0.47\textwidth]{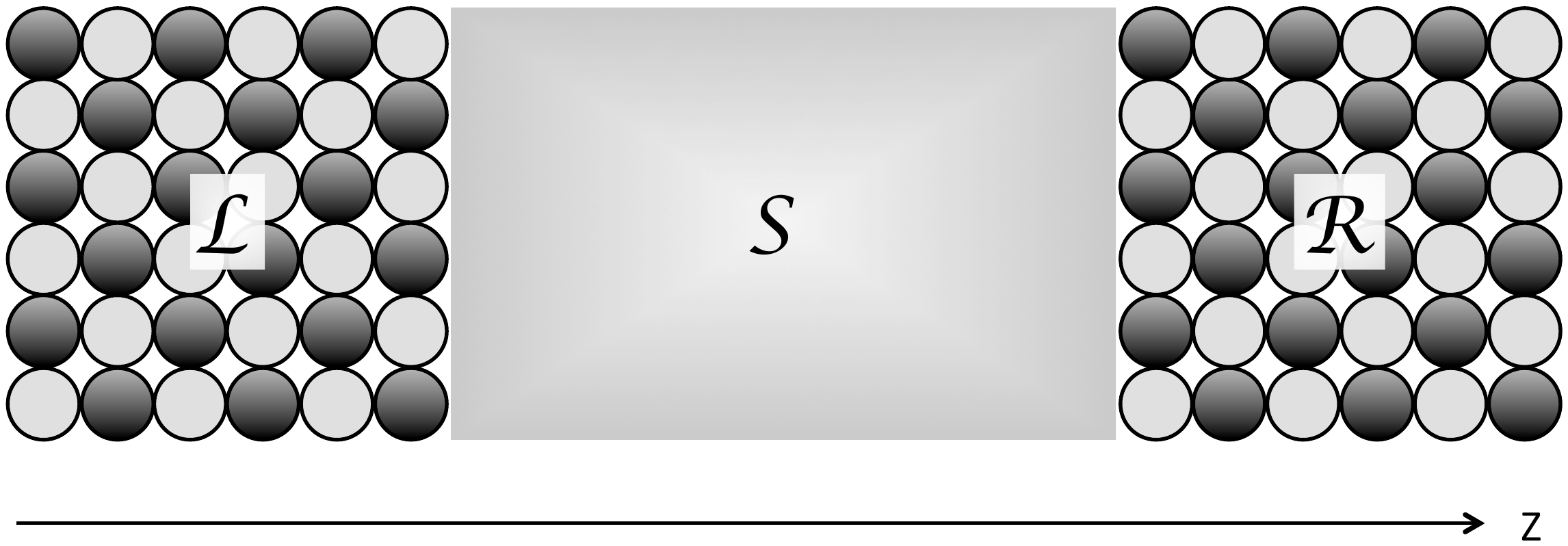}
\caption{
A sketch of a two-terminal configuration for the scattering problem. A gray scattering region ($\mathcal{S}$) is sandwiched between left ($\mathcal{L}$) and right ($\mathcal{R}$) semi-infinite leads that have translational symmetry. A current flows along the direction of the $z$-axis.}
\label{fig:geom}
\end{figure}

Central to the method is the scattering formalism \cite{Ando:prb91} for the conductance of a two-terminal device \cite{Datta:95}. The system under investigation is attached to reservoirs by semi-infinite leads that support well defined scattering states (Fig.~\ref{fig:geom}). For crystalline leads, these states are right- and left-propagating Bloch states that are incident upon the scattering region and either reflected from it or transmitted through it. The probability amplitude that the $\nu^{\rm th}$ right-propagating state incident from the left lead with spin $\sigma'$ is scattered into the $\mu^{\rm th}$ right-propagating state with spin $\sigma$ in the right lead defines the transmission matrix element $t^{\sigma\sigma'}_{\mu\nu}$. The crystal momenta and band indices of the scattering states are labelled by $\mu$ or $\nu$. Similarly, the reflection matrix $r^{\sigma\sigma'}_{\mu\nu}$ can be defined for right-propagating states that are reflected back into left-propagating states in the left lead. Denoting leads as left ($\mathcal{L}$) and right ($\mathcal{R}$) for a two-terminal system, we have two transmission matrices ${\bf t}_{\mathcal {LR}}$ and ${\bf t}_{\mathcal{RL}}$, for electrons coming from left- and right-leads, and, similarly, two reflection matrices ${\bf r}_{\mathcal {LL}}$ and ${\bf r}_{\mathcal{RR}}$. Together, they form the scattering matrix ${\bf S}$ 
\begin{equation}
{\bf S}=\begin{pmatrix}
{\bf r}_{\mathcal {LL}} & {\bf t}_{\mathcal{RL}} \\
{\bf t}_{\mathcal {LR}} & {\bf r}_{\mathcal{RR}} 
\end{pmatrix}
 \label{eq:scatt}
\end{equation}
that contains all the information needed to study a number of important physical properties of the system. The best known such property is the conductance $G$ that can be expressed according to the Landauer-B{\"u}ttiker formalism as \cite{Datta:95, Imry:02} 
\begin{equation}
  G=\frac{e^2}{h}\mathrm{Tr}\{{\bf t} {\bf t}^{\dagger}\}\;.
  \label{eq:LB}
\end{equation}
The scattering formalism is not restricted to calculations of the conductance but can provide us with useful information about spin-dynamics and spin-relaxation processes in the scattering region. In particular, it can be used to calculate the Gilbert damping parameter $\alpha$ and the spin-flip diffusion length $l_{\rm sf}$ \cite{Starikov:prl10, LiuY:prl14, LiuY:prb15}. 

The present study is based upon a first-principles tight-binding (TB) linearized-muffin-tin-orbital (LMTO) implementation of the scattering formalism. TB-LMTOs form a minimal basis set \cite{Andersen:prl84, Andersen:85, Andersen:prb86} that allows us to construct a highly efficient computational method, especially when combined with sparse-matrix techniques \cite{Amestoy:siamjm01, Amestoy:pc06}. In combination with the local spin-density approximation (LSDA) from density functional theory (DFT), it allows us to study physical systems either entirely {\it ab initio}, i.e., without introducing any free parameters or in the case of finite temperature transport, a minimal number thereof. We extend earlier work \cite{Xia:prb06, Zwierzycki:pssb08} by introducing non-collinear magnetism \cite{Wang:prb08} and spin-orbit interaction using the Pauli-Schr{\"o}dinger Hamiltonian. We generalize the wave-function matching (WMF) technique introduced by Ando \cite{Ando:prb91} to eliminate the need for a principal layer decomposition and dispense altogether with partitioning the scattering region into layers. Compared to the widely used recursive Green's function method, the interaction range in the leads and scattering region is arbitrarily long without loss of numerical stability and optimal use can be made of sparse-matrix solvers which greatly improves the computational efficiency. Not having to divide the scattering geometry into ``blocks'' or ``layers'' results in greater flexibility in applications to disorder.

We illustrate how this framework can be used to investigate spin-dependent transport in the diffusive regime and spin-relaxation phenomena using recent developments in scattering theory and spin-dynamics \cite{Brataas:prl08, *Brataas:prb11}. 
In Sect.~\ref{sec:theory} we outline the technical details of the method and illustrate it by applying it to the Ni$_{80}$Fe$_{20}$ alloy permalloy in Sect.~\ref{sec:calcs}. Some technical details of the SOC implementation with LMTOs are given in a set of appendices \ref{sec:cdappendix}. Appendix \ref{sec:limitations} contains a brief discussion of some limitations of the method as well as numerical tests about the exchange-correlation functional in the LSDA and the basis set of TB-LMTOs.

\section{Formalism}
\label{sec:theory}
To solve the scattering problem for the infinite system depicted in Fig.~\ref{fig:geom}, we need to solve the single-particle Schr\"odinger equation 
\begin{equation}
  \left({\bf H}-E\mathbf I \right) \mathbf{\Psi} = 0 ,
\label{eq:SE}  
\end{equation}
at some specified energy $E$, usually the Fermi energy. We assume that the ground state charge and spin-densities and Kohn-Sham potentials for all atoms in the system have already been calculated self-consistently.
Here, ${\bm \Psi}$ is a vector of coefficients $\Psi_i$ when the wave function $\bm{\Psi}$ is expanded in some localised orbital basis ($i \equiv {\bf R}lm\sigma$ for the MTOs we will use where ${\bf R}$ is an atom site index and $lm\sigma$ have their conventional orbital angular momentum and spin meaning; see Appendix \ref{sec:lmtosoc}). ${\bf H}$ is the Hamiltonian matrix in the localized orbital basis and a summation over $i$ is implied in \eqref{eq:SE}. Its sparsity is determined by the range of the localized orbitals which is minimal for TB-MTOs. The system in Fig.~\ref{fig:geom} is infinite so that the dimensions of the Hamiltonian $\bf H$ and unit matrix $\bf I$ in Eq.~\eqref{eq:SE} are both infinite. By applying the ``wave-function matching'' method \cite{Ando:prb91} the semi-infinite leads with full translational symmetry can be replaced with appropriate boundary conditions in the form of energy-dependent embedding potentials on the boundary layers. This reduces the problem to a finite size and results in a two-stage process for calculating the scattering matrix. In the first stage, to be discussed in Sec. \ref{sec:wire}, eigenmodes ${\bf u}_m$ of the leads are calculated by solving the Schr{\"o}dinger equation for each of the leads in turn taking translational symmetry into account. By calculating their wave vectors ${\bf k}_m$ and velocities ${\bf v}_m$, the eigenmodes can be classified as being either left-going $\mathbf{u}_m(-)$ or right-going $\mathbf{u}_m(+)$. They form a basis in which to expand any left- and right-going waves in the leads and their transformation under a layer translation in the leads is easily calculated by using a generalization of Bloch's theorem for complex ${\bf k}$ \cite{Ando:prb91}.

In the second stage, discussed in Sec.~\ref{sec:scattering}, these solutions from the first stage can be used to construct the energy-dependent boundary conditions for the Hamiltonian in the scattering region, which can be a slab of a random alloy, a single interface, a multilayered structure, a tunnel junction, a slab of thermally disordered material, etc. One then has to solve a system of linear equations (LEQs) with the original Hamiltonian modified by incorporating the boundary conditions in the role of a coefficient matrix to obtain the wave functions $\mathbf{\Psi}$ that provide all information about the scattering in the system and can be used to calculate the transmission and reflection probability amplitudes, $t_{\mu\nu}$ and $r_{\mu\nu}$, and more. As a result of choosing a localized basis to minimize the hopping range, the Hamiltonian matrix is very sparse. This can be exploited by using efficient numerical methods such as incomplete LU-factorisation (which takes into account the sparsity of the matrix) to solve the LEQs. The solution scales linearly with the extent of the scattering region in the transport direction.

Once the scattering matrix is known, we can extract the resistivity and Gilbert damping parameter as discussed in Sec.~\ref{sec:extraction}. The scattering formalism will be presented in its general form not depending on details of the underlying basis set and Hamiltonian; the most relevant aspects of the LMTOs used in the current implementation are sketched in Appendix~\ref{sec:lmtosoc}.
In Sec.~\ref{sec:formation} we discuss ways of modelling different kinds of disorder using large supercells transverse to the transport direction.

\subsection{Eigenstates of ideal leads}
\label{sec:wire}
We make use of an assumed two-dimensional (2D) translational symmetry in the plane perpendicular to the transport direction to characterise states in this and the next section with a lateral wave vector $\mathbf{k}_{\parallel}$ in the corresponding two-dimensional Brillouin zone (2D BZ). All variables therefore have an implicit dependence on $\mathbf{k}_{\parallel}$ that will be suppressed for simplicity. When we refer to the number of atoms (orbitals) in a layer, we refer to the finite number number of atoms (orbitals) in a translational unit cell.

Because the ideal leads have translational symmetry in the transport direction, they can be decomposed into an infinite number of translationally invariant layers. When the hopping range of the Hamiltonian of this system is greater than the corresponding periodicity, i.e., when hopping to layers beyond the nearest neighbouring layers is not negligible, the usual approach would be to increase the layer thickness until only hopping between neighbouring layers occurs; these are called {\em principal layers}. In general the principal layer procedure results in increased computational cost and decreased accuracy. To remedy this, we formulate the WFM method for arbitrary hopping range between layers, thus generalizing previous formulations of the WFM method \cite{Ando:prb91, Khomyakov:prb04, Xia:prb06, Zwierzycki:pssb08}.

\begin{figure}[t]
  \centering
\includegraphics[width=0.47\textwidth]{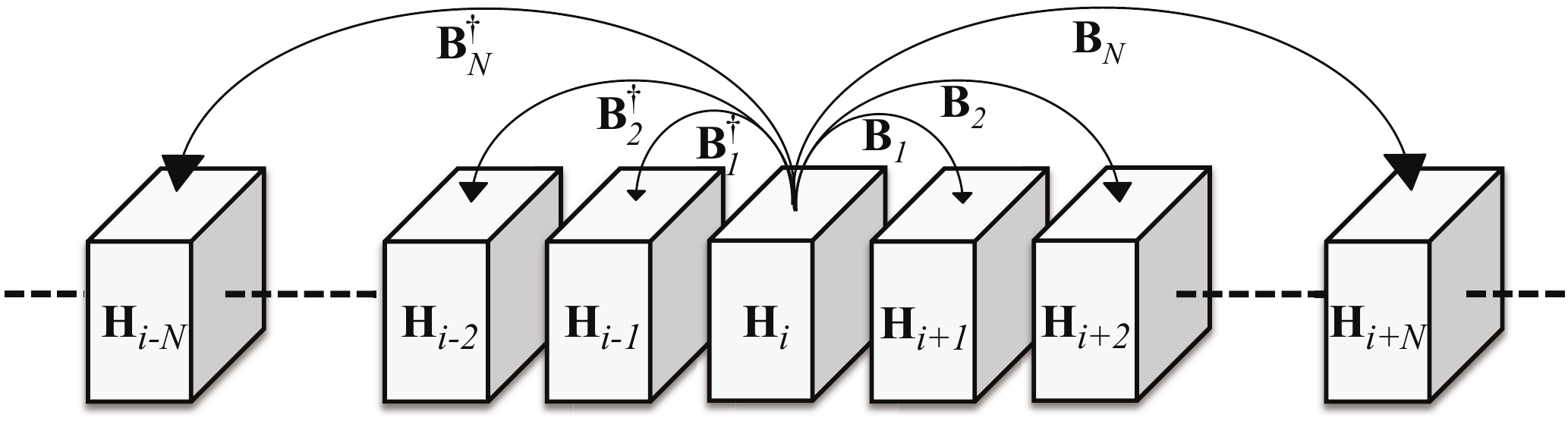}
\caption{Hamiltonian matrix of an ideal quantum wire partitioned into slices determined by the translational symmetry of the leads. ${\bf H}_i\equiv{\bf H}_{i,i}$ is the on-layer term of the Hamiltonian, ${\bf B}_l \equiv {\bf H}_{i,i+l}$ and ${\bf B}^{\dagger}_l\equiv {\bf H}_{i,i-l}$ describe hopping to the $l^{\rm th}$ and  $-l^{\rm th}$ neighbouring layers, respectively.}
\label{fig:wire}
\end{figure}

\begin{figure*}[t]
\includegraphics[scale=0.56]{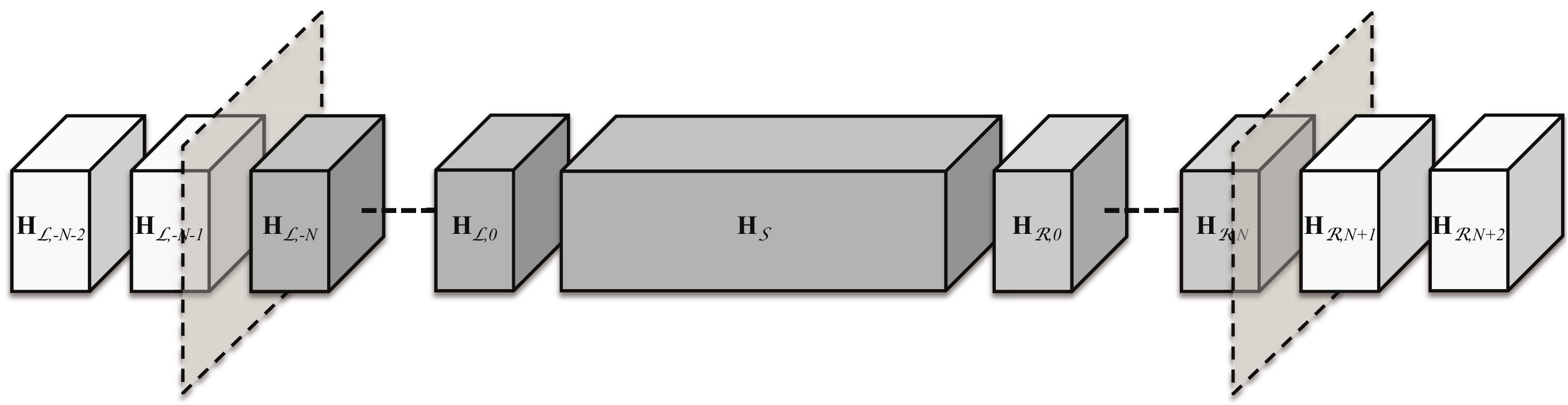}
\caption{Geometry of the finite scattering problem comprising left ($\mathcal{L}$),  and right ($\mathcal{R}$) leads sandwiching the scattering ($\mathcal{S}$) region that is augmented by a finite number $N$ of lead layers chosen to be sufficiently large so that there is no hopping from the scattering region proper to the (white) lead layers. 
}
\label{fig:geo-scatter}
\end{figure*} 

We start with an ideal wire with translational symmetry (Fig.~\ref{fig:wire}), in which every layer contains $N_O$ atom centred orbitals and is coupled to some number ($N$) of layers to its left and right. Then the Schr{\"o}dinger equation for the $i^{\rm th}$ layer is given by
\begin{equation}
\left(E {\bf I} - {\bf H}_i\right){\bf \Psi}_i 
    + \sum_{l=1}^{N} \left( 
       {\bf B}_l {\bf \Psi}_{i+l} 
    + {\bf B}^{\dagger}_l {\bf \Psi}_{i-l} \right)=0 \;, 
\label{ando:schrod}
\end{equation}
where ${\bf H}_i\equiv{\bf H}_{i,i}$ is the on-layer term of the Hamiltonian, ${\bf B}_l \equiv {\bf H}_{i,i+l}$ and ${\bf B}^{\dagger}_l\equiv {\bf H}_{i,i-l}$ describe hopping to the $l^{\rm th}$ and  $-l^{\rm th}$ neighbouring layers respectively,  and ${\bf \Psi}_{i+l}$ is the wave function on the $l^{\rm th}$ neighbouring layer. Taking into account translational symmetry in the periodic crystal, the wave function on any arbitrary layer $l$ is related to the wave function on layer $l-1$ by a generalized Bloch factor $\lambda$ as 
\begin{equation}
{\bf \Psi}_{l}=\lambda {\bf \Psi}_{l-1} \; .
\label{ando:bloch}
\end{equation}
Combining \eqref{ando:schrod} and \eqref{ando:bloch} leads to the generalised eigenvalue problem of rank $2 \times N \times N_O$  
\begin{subequations}
\begin{eqnarray}
\left(E {\bf I} - {\bf H}_0 \right){\bm \Psi}_0 + 
   \sum_{l=1}^{N-1} \big( {\bf B}_l {\bm \Psi}_l +&& {\bf B}^{\dagger}_l {\bm \Psi}_{-l} \big) +
{\bf B}^{\dagger}_N {\bf \Psi}_{-N} \nonumber \\
=&& - \lambda {\bf B}_N {\bf \Psi}_{N-1},
\end{eqnarray}
\begin{equation}
{\bf \Psi}_{l} = \lambda {\bf \Psi}_{l-1},  \;\;\;\;\;\;\;\;\;\;\;\;\;\; \forall \;\; l \in [-N+1,N-1],
\end{equation}
\label{ando:geneig}
\end{subequations}
where, without loss of generality, $i=0$ has been assumed. Non-trivial solutions of \eqref{ando:geneig} can be separated into two classes. The first class consists of $N \times N_O$ left-going waves, the second class of $N \times N_O$ right-going waves. Each class can contain both propagating Bloch waves and nonpropagating, evanescent waves. The corresponding Bloch factors are denoted by ${\bf \lambda}(+)$ and ${\bf \lambda}(-)$. Of these $N \times N_O$ solutions, only $N_O$ Bloch factors ${\bf \lambda}(+)$ correspond to the translation of right-propagating waves to a neigbouring layer, the rest of the ${\bf \lambda}(+)$ factors describe translations to more distant layers and do not provide any additional information; thus we have only $N_O$ unique translation factors among the ${\bf \lambda}(+)$. Similarly, for left-propagating waves there are only $N_O$ unique translations in the set of ${\bf \lambda}(-)$ factors. By using only the $N_O$ orbitals belonging to the $l=0$ layer with eigenvectors from \eqref{ando:geneig} corresponding to the set of unique translation factors ${\bf \lambda}(\pm)$, we construct normalised eigenvectors ${\bf u}_m(\pm)$ ($ \equiv {\bm \Psi}_{l=0,m}$ where $l$ is the layer-index and $m$ is the mode index) and use these to form the $N_O \times N_O$ matrices \cite{Ando:prb91}
\begin{equation}
  {\bf U}(\pm)=\left({\bf u}_1(\pm)\cdot\cdot\cdot{\bf u}_{N_O}(\pm)\right) \;.
  \label{eq:uvec}
\end{equation}
Any arbitrary wave function on the $l=0$ layer can then be represented as a linear combination of left- and right-propagating waves
\begin{equation}
  {\bf \Psi}={\bf \Psi}(+)+{\bf \Psi}(-) \;,
\end{equation}
and any left- or right-propagating wave can be expanded in terms of the eigenstates of the lead as
\begin{equation}
  {\bf \Psi}(\pm)={\bf U}(\pm) {\bf C}(\pm)\;,
\end{equation}
where ${\bf C}_{\mu}(\pm)$ is a vector of coefficients.
We define the $N_O\times N_O$ diagonal eigenvalue matrices by 
\begin{equation}
{\bf \Lambda}(\pm)=\delta_{nm} \lambda_m(\pm)\;.
\end{equation}
Using the Bloch condition \eqref{ando:bloch} and following Ando's original procedure \cite{Ando:prb91, Khomyakov:prb05, Xia:prb06, Zwierzycki:pssb08} we define the translation matrices 
\begin{align}
{\bf F}(\pm)&={\bf U}(\pm){\bf \Lambda}(\pm){\bf U}^{-1}(\pm) \;.
\label{eq:blochmatrices}
\end{align}
The translation of the wave function on layer $i$ over an arbitrary number of layers $l$ is then given by
\begin{align}
   {\bf \Psi}_{i+l}={\bf F}^{l}(+){\bf \Psi}_{i}(+)+{\bf F}^{l}(-){\bf \Psi}_i(-)\;,
\end{align}
allowing us to construct the full solution for the entire lead.  

\subsection{The scattering problem}
\label{sec:scattering}

The scattering region $\mathcal S$ is now inserted between the left and right leads. It is important to emphasize that we do not define a layered structure inside the scattering region. Regardless of its size or contents, the scattering region acts as one large meta-layer. The resulting problem is infinite but by making use of the translational symmetry in the leads and the solutions obtained in the previous section, the leads can be incorporated in the scattering problem in the form of boundary conditions imposed in the lead layers adjoining the scattering region. The system is partitioned as shown in Fig.~\ref{fig:geo-scatter} where the infinite scattering geometry is truncated to include only $N\!+\!1$ (translationally invariant) lead layers on the left and right in addition to the original  (disordered) scattering region where $N$ is the hopping range (in terms of number of layers) in the Hamiltonian describing the leads. 

Specifically, we assume that the $N\!+\!1$ lead layers attached to the original scattering region on the left side are indexed as $-N, \ldots, 0$. Then the wave function in the layer in the left lead with index $-(\!N\!+\!Q\!)$,  where $Q>0$ i.e., the wave function in the white layers on the left in Fig.~\ref{fig:geo-scatter}, can be related to the wave function ${\bf \Psi}_{\mathcal{L},-N}={\bf \Psi}_{\mathcal{L},-N}(+)+{\bf \Psi}_{\mathcal{L},-N}(-)$, a superposition of left- and right-propagating waves, as
\begin{align}
\label{eq:13}
&{\bf \Psi}_{\mathcal{L},-N-Q}={\bf \Psi}_{\mathcal{L},-N-Q}(+)+{\bf \Psi}_{\mathcal{L},-N-Q}(-) \nonumber\\
                  &={\bf F}_{\mathcal{L}}^{-Q}(+){\bf \Psi}_{\mathcal{L},-N}(+)
                   +{\bf F}_{\mathcal{L}}^{-Q}(-){\bf \Psi}_{\mathcal{L},-N}(-)  \\
                  &=\left[{\bf F}_{\mathcal{L}}^{-Q}(+)-{\bf F}_{\mathcal{L}}^{-Q}(-)\right]{\bf \Psi}_{\mathcal{L},-N}(+)+{\bf F}_{\mathcal{L}}^{-Q}(-){\bf \Psi}_{\mathcal{L},-N} \nonumber,
\end{align}
allowing us to express an arbitrary  ${\bf \Psi}_{\mathcal{L},-N-Q}$ in terms of ${\bf \Psi}_{\mathcal{L},-N}$ and ${\bf \Psi}_{\mathcal{L},-N}(+)$. The set of Schr{\"o}dinger equations for layers $-N,\ldots, 0$ become
\begin{widetext}
\begin{equation}
    \left(E {\bf I} - {\bf H}_{\mathcal{L},n}\right)\Psi_{\mathcal{L},n} + 
    \sum_{l=1}^{-n} {\bf B}_{\mathcal{L},l} \Psi_{\mathcal{L},n+l} + {\bf B}_{\mathcal{LS},n}  \Psi_{\mathcal{S}} 
     + \sum_{l=1}^{N} {\bf B}^{\dagger}_{\mathcal{L},l}  \Psi_{\mathcal{L},n-l}  =0 \;\;\;\;\;\;\;\;\; \forall \;\; n \in [-N,0], \;\;
\label{eq:14}
\end{equation}
where  ${\bf B}_{\mathcal{LS},n}$ is the coupling between the lead layer on the left with index $n$ (in the truncated transport geometry) to the (original) scattering region $\mathcal{S}$, ${\bf B}_{\mathcal{L},l}$ is the hopping to  the $l$-th next layer in the left lead, and $\Psi_{\mathcal{S}}$ is the wave function in the scattering region. By splitting the summation $\sum_{l=1}^{N} =\sum_{l=1}^{N+n} +\sum_{l=N+n+1}^{N} $ in the last term on the lhs and using \eqref{eq:13} to eliminate $\Psi_{\mathcal{L},n}$ ($\forall \; n \!< \!-N$), equation \eqref{eq:14} can be transformed to
%
\begin{align}
\left(E {\bf I} - {\bf H}_{\mathcal{L},n}\right)\Psi_{\mathcal{L},n} & +  \sum_{l=1}^{-n} {\bf B}_{\mathcal{L},l}\Psi_{\mathcal{L},n+l} +  {\bf B}_{\mathcal{LS},n}\Psi_{\mathcal{S}} 
 +\big(1-\delta_{N,-n}\big)\sum_{l=1}^{N+n} {\bf B}^{\dagger}_{\mathcal{L},l} \Psi_{\mathcal{L},n-l} 
 +\sum_{l=N+n+1}^{N} {\bf B}^{\dagger}_{\mathcal{L},l} {\bf F}_{\mathcal{L}}^{-l+N+n}(-){\bf \Psi}_{\mathcal{L},-N} 
\nonumber\\
&  =  - \sum_{l=N+n+1}^{N} {\bf B}^{\dagger}_{\mathcal{L},l}  \Big[{\bf F}_{\mathcal{L}}^{-l+N+n}(+) - {\bf F}_{\mathcal{L}}^{-l+N+n}(-)\Big] {\bf \Psi}_{\mathcal{L},-N} (+) \hskip .14\textwidth \forall n \in [-N,0]\;.
\label{eq:leftemb}
\end{align}
%
This effectively acts as a boundary condition for the left-hand side of \eqref{eq:leftemb} and removes a direct dependence on the wave functions to the left of the $-N^{\rm th}$ layer.

Injection of electrons from the left electrode leads to only right-propagating waves on the right-hand side of the scattering region and the wave function in the layer with index $N+Q$ can be related to the wave function in the rightmost layer ($N$) within the truncated transport geometry as
\begin{align}
  {\bf \Psi}_{\mathcal{R},N+Q}={\bf F}_{\mathcal{R}}^{Q}(+) {\bf \Psi}_{\mathcal{R},N}
\end{align}
allowing ${\bf \Psi}_{\mathcal{R},N+Q}$ to be eliminated from the Schr{\"o}dinger equation
%
\begin{align}
&  \left(E {\bf I} - {\bf H}_{\mathcal{R},n}\right){\bf \Psi}_{\mathcal{R},n} + \sum_{l=1}^{n} {\bf B}^{\dagger}_{\mathcal{R},l}{\bf \Psi}_{\mathcal{R},n-l} +  
{\bf B}_{\mathcal{RS},n}^{\dagger}{\bf \Psi}_{\mathcal{S}}   +  \big(1-\delta_{N,n}\big)\sum_{l=1}^{N-n}  {\bf B}_{\mathcal{R},l} {\bf \Psi}_{\mathcal{R},n+l} + \!\!\!\! \sum_{l=N-n+1}^N \!\!\!\! {\bf B}_{\mathcal{R},l} {\bf F}_{\mathcal{R}}^{l-N+n}(+){\bf \Psi}_{\mathcal{R},N} =0 \nonumber \\
& \hskip .8\textwidth \forall n \in [0,N] \;,
\label{eq:rightemb}
\end{align}
\end{widetext}
where ${\bf B}^{\dagger}_{\mathcal{RS},n}$ is the coupling between the scattering region and the lead layer with index $n$ (in the truncated transport geometry), and ${\bf B}_{\mathcal{R},l}$ is the hopping to the $l$-th next layer in the right lead.

Combining the sets of equations (\ref{eq:leftemb}) and (\ref{eq:rightemb}) with the Schr{\"o}dinger equation for the scattering region
\begin{align}
  \left(E {\bf I} - {\bf H}_{\mathcal{S}}\right){\bf \Psi}_{\mathcal{S}}+\sum_{l=0}^{N-1}\left[ {\bf B}_{\mathcal{LS},l}^{\dagger} {\bf \Psi}_{\mathcal{L},-l} +{\bf B}_{\mathcal{RS},l} {\bf \Psi}_{\mathcal{R},l} \right] = 0 \;
\end{align}
results in a set of inhomogeneous LEQs. By assuming that ${\bf \Psi}_{\mathcal{L},-N}(+)={\bf U}_{\mathcal{L}}(+)$ and solving the LEQs with multiple right-hand sides in one go, we can obtain the wave functions in the system for electrons that are injected in all possible propagating modes of the left lead (i.e., modes with $|\lambda|=1$).

Bloch states incident from the left and propagating to the right are scattered by the breaking of translation symmetry into left-going states on the left-hand side and right-going states on the right-hand side as
\begin{subequations}
\begin{align}
  {\bf \Psi}_{\mathcal{L},-N}(-)&={\bf U}_{\mathcal{L}}(-) \; {\bf \tilde r} \;, \\
  {\bf \Psi}_{\mathcal{R},N}(+)&={\bf \Psi}_{\mathcal{R},N} ={\bf U}_{\mathcal{R}}(+) \; {\bf \tilde t} \;.
\end{align}
\end{subequations}
Once we know the set of wave functions ${\bf \Psi}$ for all incoming states from the left lead we can calculate the elements of the matrices ${\bf \tilde{r}}= \tilde{r}_{\mu\nu}$ with dimension $M_{\mathcal{L}}\times M_{\mathcal{L}}$, where $M_{\mathcal{L}}$ is the number of propagating modes in the left lead, and ${\bf \tilde{t}}= \tilde{t}_{\mu\nu}$ with dimensions $M_{\mathcal{R}} \times M_{\mathcal{L}}$
\begin{subequations}
\begin{align}
{\bf \tilde r}&= {\bf U}_{\mathcal{L}}^{-1}(-) \Big[{\bf \Psi}_{\mathcal{L},-N}-{\bf U}_{\mathcal{L}}(+)\Big]\;, \\
{\bf \tilde t}&= {\bf U}_{\mathcal{R}}^{-1}(+) {\bf \Psi}_{\mathcal{R},N} \;.
\end{align}
\end{subequations}
The elements of the physical reflection and transmission probability amplitude matrices can be found by normalising with respect to the currents
\begin{equation}
  r_{\mu\nu}=\sqrt{\frac{v_{\mathcal{L},\mu}(-)}{v_{\mathcal{L},\nu}(+)}} \; {\tilde r}_{\mu\nu} \;\; ; \;\;
  t_{\mu\nu}=\sqrt{\frac{v_{\mathcal{R},\mu}(+)}{v_{\mathcal{L},\nu}(+)}} \; {\tilde t}_{\mu\nu} \,
\end{equation}
where $v_{\mathcal{R,L}}(\pm)$ are the group velocities of the eigenmodes in the left and right leads, which are determined using the expressions derived in Appendix~\ref{sec:veloc}
\begin{equation}
v_{\nu}(\pm)= \frac{2a}{\hbar} \sum_{n=1}^N n \:\mathrm{Im}\Big[ \lambda^n_{\nu}(\pm) {\bf u}_{\nu}^{\dagger}(\pm) {\bf B}_n {\bf u}_{\nu}(\pm) \Big] \;.
\label{ando:veloc}
\end{equation}

When SOC is included, spin is no longer a good quantum number and separating the equation of motion for different spinors is not possible. Nevertheless, it will be convenient for the purposes of analysing our results to decompose the matrices $r_{\mu\nu}$ and $t_{\mu\nu}$ into the spin-projections $r_{\mu\nu}^{\sigma\sigma\prime}$ and $t_{\mu\nu}^{\sigma\sigma\prime}$. This is discussed in Appendix~\ref{sec:decompose}. 

When only nearest neighbour hopping is allowed ($N=1$), the expressions in sections~\ref{sec:wire} and~\ref{sec:scattering} reduce to expressions known from earlier work~\cite{Ando:prb91, Xia:prb06, Khomyakov:prb05, Zwierzycki:pssb08}. For arbitrary values of $N$ they represent a generalised WFM technique that is similar to the widely used recursive Green's function method. The advantage is that proper treatment of sparsity of the resulting LEQs allows us to use efficient sparse-matrix LEQ solvers which drastically improves computational efficiency. Additionally, departure from the recursive Green's function method allows us to describe the scattering region without introducing ``blocks" or ``layers''. This eliminates numerical issues and simplifies application of the method in complicated, incommensurable systems. The equivalence of the WFM method with the Kubo-Greenwood formalism in the linear-response regime was shown earlier \cite{Khomyakov:prb05}.

\subsection{Extracting material-specific parameters from the scattering matrix}
\label{sec:extraction}
Once we know the scattering matrix \eqref{eq:scatt} consisting of the $r_{\mu\nu}^{\sigma\sigma\prime}$ and $t_{\mu\nu}^{\sigma\sigma\prime}$ matrices calculated from the left and right-hand sides, we can use this to extract various bulk (and interface) parameters currently of interest in the field of spintronics. We focus here on the bulk resistivity and Gilbert damping of the important Ni$_{80}$Fe$_{20}$ ferromagnetic alloy, permalloy as illustrative examples.

\subsubsection{Resistivity}
\label{sec:res}
The total resistance of a diffusive conductor (e.g. alloy) of length $L$ sandwiched between two identical ideal (ballistic) leads can be expressed as
\begin{equation}
  1/G = 1/G_{\rm Sh} +R \;,
\end{equation}
where $G$ is the total conductance of the system and $G_{\rm Sh}= \left( 2 e^2/h \right) N$ is the Sharvin conductance of each lead with $N$ conductance channels per spin. $R$, the resistance of the scattering region corrected for the finite conductance of the ballistic leads~\cite{Schep:prb97, Xia:prb06}, has two contributions
\begin{equation}
	R(L)=2 R_i + R_b(L) \; ,
\end{equation}
where $R_i$ is the resistance of a single alloy$|$lead interface, and $R_b(L)$ is the bulk resistance of an alloy layer of thickness $L$.  For a sufficiently thick alloy layer, Ohmic behaviour is recovered when $R_b(L)\approx \rho L$, where $\rho$ is the bulk resistivity. 

In materials whose SOC is weak, the transport of electrons is found to be well described by considering currents of spin-up and spin-down electrons separately. For a stack of materials comprising ferromagnetic (FM) and nonmagnetic (NM) metals, the resistance of each spin channel is obtained by adding resistances in series, the two spin channels are then added in parallel according to the ``two-current series-resistor'' (2CSR) model \cite{Zhang:jap91, Lee:jmmm93, Valet:prb93}. When a ferromagnetic alloy is sandwiched between nonmagnetic leads, each spin species sees two spin-dependent interface resistances $R_i^{\sigma}$ and a spin-dependent bulk term: $R^{\sigma}(L) = 2 R_i^{\sigma}+\rho^{\sigma} L $. The total resistance that results from adding these terms in parallel can be written as
\begin{align}
  R(L)=&\frac{2 R_i \left(\beta ^2-2 \beta  \gamma +1\right)}{1-\gamma ^2} +\rho L + \nonumber \\   
  &+\frac{4 R_i^2\left(\beta^2-1\right)  {(\beta -\gamma)}^2}
  {\left(\gamma^2-1\right) \big[\left(\gamma ^2-1\right) \rho L  +\left(\beta^2-1\right) 2 R_i \big]} \;,
  \label{eq:FMresistance}
\end{align} 
in terms of the total interface resistance $R_i$ given by
\begin{equation}
	\frac{1}{R_i}=\frac{1}{R_i^{\uparrow}}+\frac{1}{R_i^{\downarrow}} ,
\end{equation}
the corresponding interface spin asymmetry $\gamma =(R_i^{\downarrow} - R_i^{\uparrow})/(R_i^{\uparrow}+R_i^{\downarrow})$, the total resistivity of the alloy $\rho$ given by 
\begin{equation}
	\frac{1}{\rho}=\frac{1}{\rho_i^{\uparrow}}+\frac{1}{\rho_i^{\downarrow}} ,
\end{equation}
and the corresponding bulk spin asymmetry $\beta = (\rho^{\downarrow} - \rho^{\uparrow})/(\rho^{\downarrow} + \rho^{\uparrow})$. 

When the SOC can no longer be considered weak, eight parameters are used to describe the resistance of a diffusive NM$|$FM$|$NM system \cite{Bass:jmmm16}. Two parameters are required to describe the NM metal, a resistivity $\rho_{\rm NM}$ and a spin-flip diffusion length $l_{\rm NM}$. Three parameter are required to describe the FM metal: a resistivity $\rho_{\rm FM}^{\sigma}$ for each spin channel as well as a spin-flip diffusion length $l_{\rm FM}$. And three parameters are required to describe the interface; an interface resistance $R_i^{\sigma}$ for each spin channel and an interface spin-flip scattering parameter $\delta$ (also called the spin memory loss parameter) \cite{Fert:prb96b, Eid:prb02, Bass:jpcm07}. The nontrivial evaluation of all of these parameters would go beyond the present task of illustrating the use of the scattering formalism and will be the subject of a separate publication \cite{Gupta:tbp18}. For the purpose of extracting a resistivity from a series of calculations of $R(L)$ we will use \eqref{eq:FMresistance} in the form
\begin{equation}
R(L)=a+\rho L +b/(\rho L+c)	\;.   \label{eq:FMres_simple}
\end{equation}
For sufficiently thick slabs where the third term in (\ref{eq:FMres_simple}) vanishes, $\rho$ will be extracted from the slope of $R(L)$. Otherwise all 4 independent parameters will be used to perform the fit. Both approaches will be examined in Sec.~\ref{res:resSOC}.

\subsubsection{Gilbert damping}
\label{sec:gd}
The magnetisation dynamics of ferromagnets is commonly described using the phenomenological Landau-Lifshitz-Gilbert equation
\begin{equation}
  \frac{d{\bf M}}{dt} =  - \gamma {\bf M} \times {\bf H}_{\rm eff} + {\bf M} \times 
  \left[ \frac{\widetilde G({\bf M})}  {\gamma M_s^2} \cdot \frac{d{\bf M}}{dt} \right] \:,
\label{eqn:LLG}
\end{equation}
where $M_s=|{\bf M}|$ is the saturation magnetisation, ${\widetilde G({\bf M})}$ is the Gilbert damping parameter (that is in general a tensor) and the gyromagnetic ratio $\gamma=g \mu_B/\hbar$ is expressed in terms of the Bohr magneton $\mu_B$ and the Land\'{e} $g$ factor, which is approximately 2 for itinerant ferromagnets. For a monodomain ferromagnetic layer sandwiched between nonmagnetic leads, NM$|$FM$|$NM, the energy dissipation due to Gilbert damping is
\begin{align}
  \frac{dE}{dt} &= \int_V d^3r \frac{d}{dt} \left( {\bf H}_{\rm eff} \cdot {\bf M} \right)\nonumber \\ 
   &= \int_V d^3r \, {\bf H}_{\rm eff} \cdot \frac{d{\bf M}}{dt}   
   = \frac{1}{\gamma^2} \frac{d{\bf m}}{dt} \cdot {\widetilde G({\bf M})} \cdot \frac{d{\bf m}}{dt} 
\label{eqn:dEdt}  
\end{align}
where ${\bf m}={\bf M}/M_s$ is the unit vector of the magnetisation direction for the macrospin mode. By equating this energy loss to the energy flow into the leads~\cite{Avron:prl01, *Moskalets:prb02a, *Moskalets:prb02b} associated with ``spin-pumping''~\cite{Tserkovnyak:prl02a, *Tserkovnyak:prb02b},
\begin{equation}
   I_E^{\rm Pump} = \frac{\hbar}{4\pi} \text{Tr} \left\{ {\frac{d{\bf S}}{dt} \frac{d{\bf S}^\dag}{dt}} \right\} = \frac{\hbar}{4\pi} \text{Tr} \left\{ \frac{d{\bf S}}{d{\bf m}} \frac{d{\bf m}}{dt} \frac{d{\bf S}^\dag}{d{\bf m}} \frac{d{\bf m}}{dt} \right\}\:,
\label{eqn:IEpump}     
\end{equation}
the elements of the tensor ${\widetilde G}$ were expressed in terms of the scattering matrix~\cite{Brataas:prl08, *Brataas:prb11}
\begin{equation}
  \widetilde G_{ij} ({\bf m}) = \frac{\gamma ^2 \hbar}{4\pi}{\mathop{\rm Re}\nolimits} \left\{ 
\text{Tr} \left[ \frac{\partial {\bf S}}{\partial m_i}\frac{\partial {\bf S}^\dag}{\partial m_j} \right] \right\} \:. 
\label{eqn:dampeq}
\end{equation}
Physically, energy is transferred from the slowly varying spin degrees of freedom to the electronic orbital degrees of freedom where it is rapidly lost to the lattice (phonon degrees of freedom). Our calculations focus on the role of elastic scattering as the rate-limiting first step.

To calculate the Gilbert damping tensor $\widetilde G_{ij}$ using \eqref{eqn:dampeq}, we need to numerically differentiate the scattering matrices with respect to the magnetisation orientation ${\bf m}$. Expressing this orientation in spherical coordinates $(\theta ,\phi)$ with the polar angle $\theta=0$ corresponding to the equilibrium magnetization direction ${\bf m}$, we vary the magnetisation direction about $\theta=0$ to calculate the $2\times 2$  damping tensor in a plane orthogonal to ${\bf m}$. Specifically, $\partial\mathbf S/\partial m_i$ in \eqref{eqn:dampeq} can be replaced by $\partial\mathbf S/\partial e_i$, where $e_i$ are components of the Cartesian basis vectors in the plane orthogonal to ${\bf m}$ and $\phi_0$ defines the orientation of the coordinate system in this plane. Then the derivatives of the scattering matrix can be approximated as
\begin{subequations}
\begin{align}
\frac{\partial S}{\partial e_1} &\approx \,\, \frac{S(\Delta\theta,\phi_0)-S(\Delta\theta,\phi_0+\pi)}{2\Delta \theta} \;,\\
\frac{\partial S}{\partial e_2} &\approx \,\, \frac{S(\Delta\theta,\phi_0+\pi/2)-S(\Delta\theta,\phi_0+3\pi/2)}{2\Delta \theta}\;, 
\end{align}
\label{eq:numdiff}
\end{subequations}
where $\Delta\theta$ is a small variation of the polar angle. Substitution of \eqref{eq:numdiff} into \eqref{eqn:dampeq} yields four elements of the $3\times 3$ damping tensor for any particular orientation ${\bf m}$ of the magnetization. For cubic substitutional alloys the damping can be assumed to be isotropic (see Appendix~\ref{sec:isotropic}), so we limit ourselves to differentiating about a single orientation ${\bf m}$ and our primary interest will be in the diagonal elements of $\widetilde G = \widetilde G_{ii}$. When the damping is enhanced by  FM$|$NM interfaces, \cite{Mizukami:jmmm01, Mizukami:jjap01, Tserkovnyak:prl02a, Zwierzycki:prb05} the total damping of a ferromagnetic slab of thickness $L$ sandwiched between leads can be written
$  {\widetilde G}(L)={\widetilde G}_{\rm if} + {\widetilde G}_b(L) $
where ${\widetilde G}_{\rm if}$ is the interface damping enhancement and we express the bulk damping in terms of the dimensionless Gilbert damping parameter $\alpha$ as
\begin{equation}
  {\widetilde G_b}(L)=\alpha \gamma M_s(L) = \alpha \gamma \mu_s A L  \;, \label{eq:dampfit}
\end{equation}
where $\mu_s$ is the magnetisation density and $A$ is the cross section. 

\subsection{Modelling disorder}
\label{sec:formation}

The structures used in spintronics studies are typically stacked layers of magnetic and nonmagnetic materials that exhibit various types of disorder. The magnetic materials themselves are frequently magnetic alloys like permalloy that are intrinsically ``chemically'' disordered and are chosen to have desirable magnetic properties. Even when two materials (like Fe and Cr) have the same crystal structure (bcc) and are closely lattice matched, it is not possible to exclude intermixing at an interface and becomes desirable to be able to model it.

Other important material combinations such as permalloy and Pt have a large lattice mismatch. For thin layers, this can be accommodated by straining either or both layers (pseudomorphic growth) but above a critical thickness these will relax to their preferred structures with or without the formation of misfit dislocations. Since fully relaxed interface structures can only be modelled using lateral supercells \cite{LiuY:prl14, WangL:prl16}, we apply the supercell approach to all forms of disorder studied in this work.

Lastly, many experiments are performed at room and elevated temperatures making it desirable to take into account temperature induced lattice and spin disorder. Our approach will be to model such thermal disorder in large lateral supercells. By doing so we will be able to make contact with a large body of experiments that have been interpreted with phenomenological models \cite{Valet:prb93} that assume a diffusive transport regime.

\subsubsection{Chemical disorder (random alloys)}
\label{sec:chem_disorder}
To study bulk alloys and interface mixing, we can calculate atomic sphere (AS) potentials self-consistently using the coherent-potential approximation (CPA) implemented with MTOs~\cite{Kudrnovsky:prb90, Kudrnovsky:cjp99} or with periodic supercells \cite{Andersen:prl84, Andersen:85, Andersen:prb86}. Transport calculations are performed with large lateral $M\times N$ supercells in which atomic sites are randomly populated with AS potentials subject to the constraint imposed by the stoichiometry of the targeted experimental system. This can be done either by enforcing the desired stoichiometry layer by layer or globally. 

\begin{figure}[b]
  \centering
\includegraphics[width=0.45\textwidth]{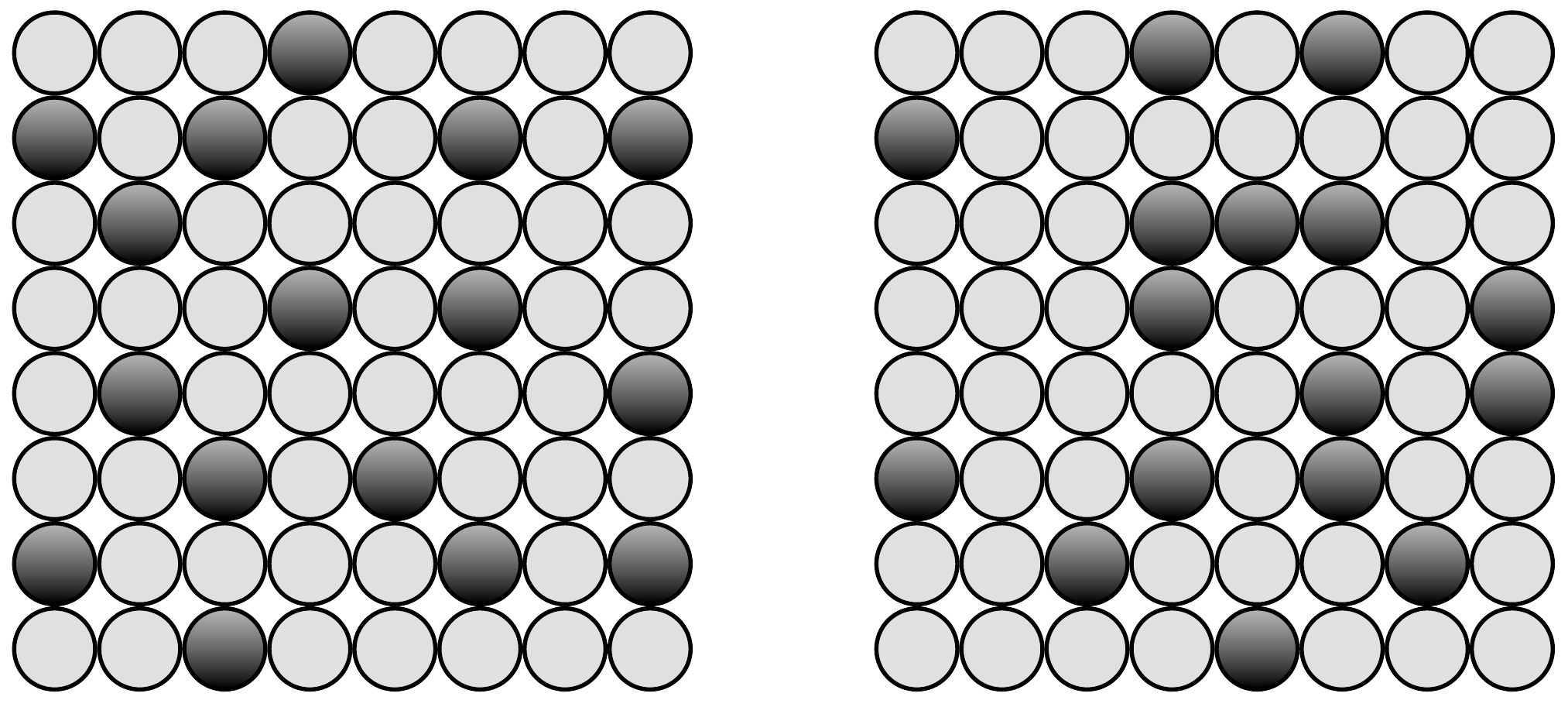}
\caption{Illustration of the configuration for a supercell with chemical disorder. Two arbitrary layers in an $8\times8$ supercell are shown for an  $A_{25}B_{75}$ alloy. Atoms of type $A$ are black and atoms of type $B$ are grey.}
\label{fig:chemdis}
\end{figure}

For example, to simulate a (001) simple cubic $A_{25}B_{75}$ random alloy using an $8\times8$ lateral supercell, we can randomly assign 16 out of the 64 sites to $A$ atoms and the rest to $B$ atoms maintaining the 25:75 stoichiometry in every layer as sketched in Fig.~\ref{fig:chemdis}. In the second case, we assign elements $A$ and $B$ randomly throughout the complete slab of material that is chemically  disordered. In both cases, configuration averaging is carried out by repeating the scattering calculations for a number of different realisations of random disorder. 

\subsubsection{Positional (thermal) disorder}
\label{sec:thermal_disorder}

\begin{figure}[t]
  \centering
 \includegraphics[width=0.45\textwidth]{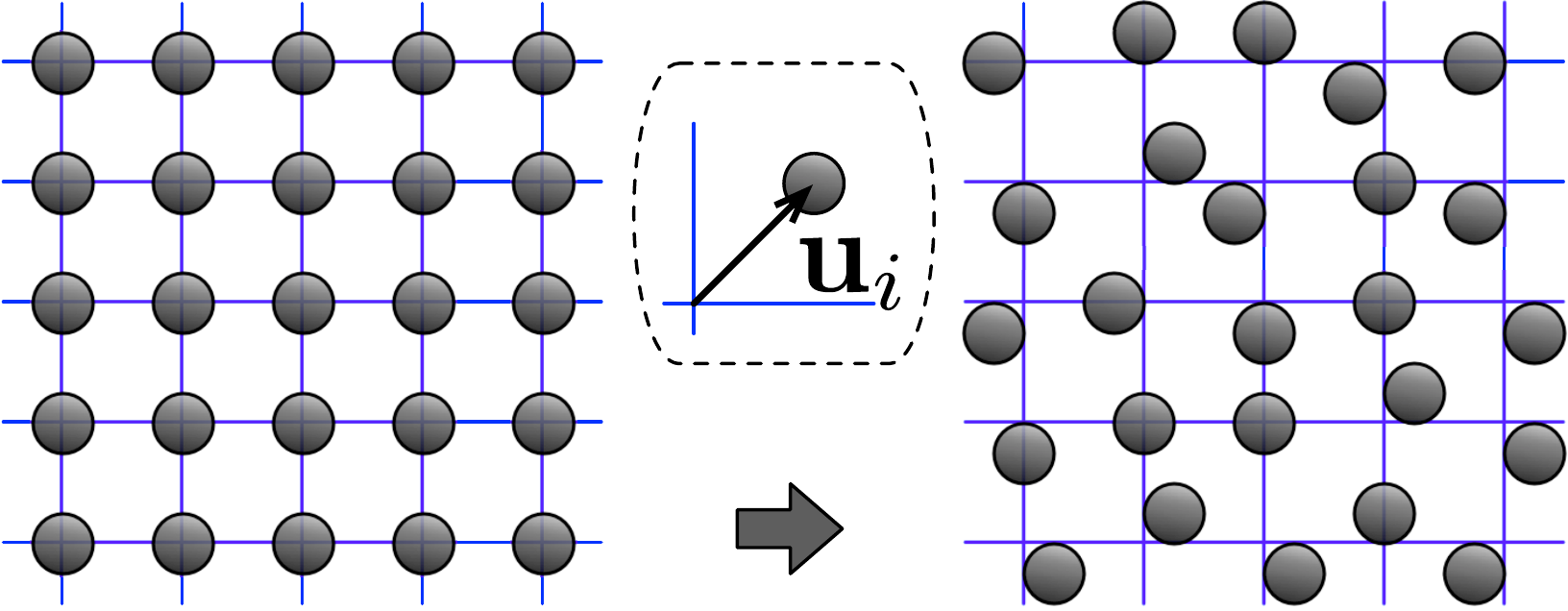}
\caption{Schematic of frozen thermal lattice disorder. Atoms (gray balls) on an ideal lattice (left panel) are displaced from their equilibrium positions by random vectors ${\bf u}_i$ to form a static configuration (right panel) for the electronic scattering calculation.}
\label{fig:displacement}
\end{figure}

Thermal lattice disorder (or other kinds of positional disorder) can be modelled in lateral supercells by displacing atoms in the scattering region from their equilibrium positions, denoted ${\bf R}_i$, by a randomised displacement vector ${\bf u}_i$ for each atomic site resulting in the new set of atomic coordinates ${\bf \widehat{R}}_i={\bf R}_i+{\bf u}_i$ (Fig.~\ref{fig:displacement}). The scattering matrices are then calculated for a number of such disordered configurations and the results averaged. The main physical approximation which is invoked is the adiabatic approximation. Though formally problematic for a metal with a gapless spectrum, within the framework of the lowest order variational approximation (LOVA) to the Boltzmann equation, the adiabatic approximation is found to describe the thermal and electrical transport properties of transition metals very well \cite{Savrasov:prb96b}.

Different approaches to generating $\mathbf{u}_i$ are possible, ranging from ab-initio molecular dynamics, through first-principles lattice dynamics \cite{LiuY:prb15}, to parameterized Gaussian disorder \cite{LiuY:prb11, LiuY:prb15}. The latter and simplest approach, that is adopted here, is based upon the harmonic approximation whereby the energy cost of displacing atoms is quadratic in their displacements. Components of ${\bf u}_i$ are then distributed normally with a root-mean-square (rms) deviation $\Delta$ that can be chosen in different ways. It can be related to the temperature and extracted from experiment within the Debye model. Or it can be chosen to reproduce an experimental temperature dependent resistivity. As the particular method of generating displacements does not affect the implementation of the transport method, we will not concern ourselves overly with the relationship between the displacements $\{{\bf u}_i\}$ and the temperature in this paper.

\subsubsection{Non-collinear configurations and magnetic disorder}
\label{sec:noncollinear}

\begin{figure}[b]
  \centering
\includegraphics[width=0.45\textwidth]{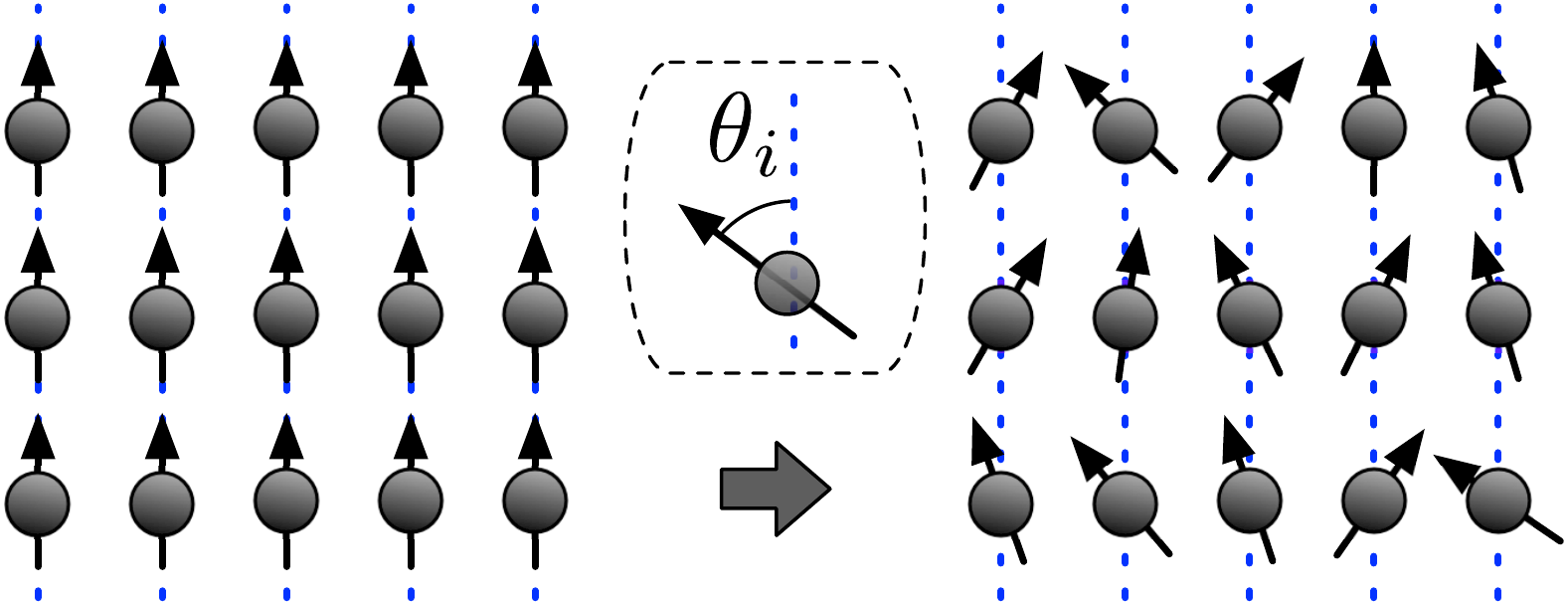}
\caption{Schematic of frozen thermal spin disorder. Atomic magnetic moments (arrows) of a ferromagnetically ordered system (left panel) are tilted by random polar angles $\theta_i$ (and azimuthal angles $\phi_i$, not shown) to form a static spin-disordered configuration (right panel) for the electronic scattering calculation.}
\label{fig:spindis}
\end{figure}

For collinear ferromagnets (or antiferromagnets), thermal spin disorder can be modelled by rotating magnetic moments in the scattering region away from their equilibrium orientations \cite{LiuY:prb11} (Fig.~\ref{fig:spindis}). To lowest order in the polar angles describing this orientation, the energy varies quadratically and temperature-induced spin disorder can be modelled with Gaussian disorder. In a magnetic domain wall separating domains in which the magnetization is collinear, the magnetization rotates continuously from one preferred orientation to the other. Addressing both these problems, spin disorder and domain walls, requires an implementation of the scattering formalism whereby atoms on different sites can have differently oriented magnetization directions. This is simply achieved in the atomic spheres approximation (ASA) \cite{Sandratskii:jpf86, Kubler:jpf88, Sandratskii:ap98}.

We assume that the spin quantisation axis $\sigma_z$ of the collinear system and the spatial $z$-axis are collinear with the direction of transport. We then rotate the local magnetic moment (exchange potential) on an arbitrary site (atomic sphere) by rotating all spin-dependent atomic parameters, which are $2\times 2$ operators in spin-space [such as the potential function $\mathbf P^{\alpha}$ \eqref{eq:potentialfunc} or SOC parameters \eqref{eq:SOCparameters}], using the rotation operator 
\begin{equation}
  {\bf \widehat R}_{si}=\exp\Big(\frac{i \sigma_y \theta_i}{2}  \Big) \exp\Big(\frac{i \sigma_z \phi_i}{2}\Big)\;,
\label{eq:rotop}
\end{equation}
where $\theta_i$ and $\phi_i$ are polar and azimuthal angles in the AS on site ${\bf \widehat R}_i$, while leaving spin-independent operators like the structure constant matrix unchanged. The LMTO Hamiltonian (\ref{eq:socham}) can then be constructed in the usual way using matrix operations on the modified operators. 

By using a suitable distribution of spinor-rotation angles, we can simulate thermal disorder or ordered structures like domain walls \cite{LiuY:prb11, YuanZ:prl12, YuanZ:prl14, YuanZ:prb16}. For thermal disorder, we can choose $\phi_i$ to be random while assuming a Gaussian distribution for $\theta_i$ with an angular rms deviation $\Delta\Theta$ related to the temperature using some model e.g. to reproduce an experimental temperature dependent resistivity \cite{LiuY:prb11} or magnetization \cite{LiuY:prb15}. Alternatively, we can calculate the interatomic exchange interactions from first-principles and use these to determine the magnon dispersion relations. By occupying the magnon modes for some chosen temperature, random sets of $\phi_i$ and  $\theta_i$ can be generated and used as input to a scattering calculation in a frozen-magnon approximation \cite{LiuY:prb15}. Details of how $\{\theta_i,\phi_i\}$ depend on temperature fall outside the scope of this paper.

Although the above scheme is only applied to magnetization fluctuations about the global quantization axis in this paper, we have applied it to nontrivial noncollinear magnetizations such as spin spirals and magnetic domain walls in references [\onlinecite{YuanZ:prl12, YuanZ:prl14, YuanZ:prb16}]. By explicitly taking the spatially varying magnetization into account, we calculated the domain wall resistance \cite{YuanZ:prl12}, the enhancement of the Gilbert damping by noncollinearity \cite{YuanZ:prl14}, and the nonadiabatic STT parameter \cite{YuanZ:prb16} in domain walls with different profiles. It could equally well be applied to study transport properties in spin glasses \cite{Niimi:prl15} or amorphous magnets \cite{Wesenberg:natp17} where the Gilbert damping is generally an anisotropic tensor depending on the symmetry \cite{Hals:prb14, Tserkovnyak:prb17}, as demonstrated for magnetic domain walls \cite{YuanZ:prl14}.

\section{Calculations}
\label{sec:calcs}

The scattering calculations are carried out in two distinct steps. In the first step semirelativistic \cite{Koelling:jpc77} AS potentials are calculated self-consistently for the atoms in the structure we are interested in, starting with the calculation of ``bulk'' potentials for the left and right leads. AS potentials can be calculated self-consistently for the scattering region using the surface Green's function (SGF) method \cite{Turek:97} or a supercell approach with a conventional ``bulk'' band structure code \cite{Andersen:prl84, Andersen:85, Andersen:prb86}. For substitutional random alloys, this is done very efficiently by combining the SGF method with the CPA \cite{Turek:97}. 
In the second step, the WFM method outlined in Sect.~\ref{sec:theory} is used to calculate the scattering matrix for the fully relativistic Pauli-Schr{\"o}dinger Hamiltonian using a TB-MTO basis; for details see Appendix~\ref{sec:lmtosoc}. In this step the scattering states in the left and right leads are first determined following Sect.~\ref{sec:wire} and then the scattering problem is solved according to Sect.~\ref{sec:scattering}.

Although the calculations are entirely {\it ab initio} in the sense that the computational scheme does not contain any free parameters, the results do depend on the numerical implementation that is necessarily approximate. In this section we discuss a number of relevant issues and illustrate some potential difficulties for a Cu$|$Py$|$Cu system consisting of a length $L$ of permalloy sandwiched between copper leads. Both fcc materials are chosen to have a $(111)$ orientation in the transport direction. The lattice constant of permalloy is taken to be $a_{\rm Py}= 3.5412 \mathring{A}$ according to Vegard's law. This is slightly smaller than the experimental lattice constant of $\rm Cu$, $a_{\rm Cu}=3.614 \mathring{A}$. As we will be interested only in the bulk properties of the alloy, we will choose the lattice constant of the copper leads to be equal to that of permalloy and ignore the (small) modifications introduced into the electronic structure of $\rm Cu$ that may result. 

Should the lattice mismatch be important, as is the case when modelling the interface properties of an A$|$B interface between materials A and B, the lattice constant ratio $a_A/a_B$ can be approximated by the ratio of two integers $N_A$ and $N_B$ such that $N_A a_A \sim N_B a_B $. When the A and B lattices are chosen to be aligned, this can result in unfavourably large values of $N_A$ and $N_B$. By dropping the alignment condition, more flexibility can be achieved by searching for lattice vectors in each lattice whose lengths match; in general this will require rotating the lattices with respect to one another. This approach made it possible to study fully relaxed interfaces between permalloy and the nonmagnetic materials Cu, Pd, Ta and Pt \cite{LiuY:prl14, WangL:prl16}.

\subsection{Modelling diffusive transport with supercells}
\label{sec:res}

The (lateral) supercell approach allows us to flexibly model interfaces between materials with different crystal structures and lattice constants by imposing a degree of lattice periodicity in order to be able to use Bloch's theorem. A substitutional alloy, or a crystalline material at finite temperature has, however, no translational symmetry and the supercell approach is formally only correct in the thermodynamic limit. Just as practical experience has shown that periodic boundary conditions can be very effectively used to model symmetry breaking surfaces, interfaces, impurities, etc. with very small supercells, we will see that we can model diffusive transport with lateral supercells of very modest size.

\subsubsection{$\Gamma$-point calculations with large lateral supercells}
\label{sec:resGamma}

\begin{figure}[b]
\centering
\includegraphics[width=0.47\textwidth]{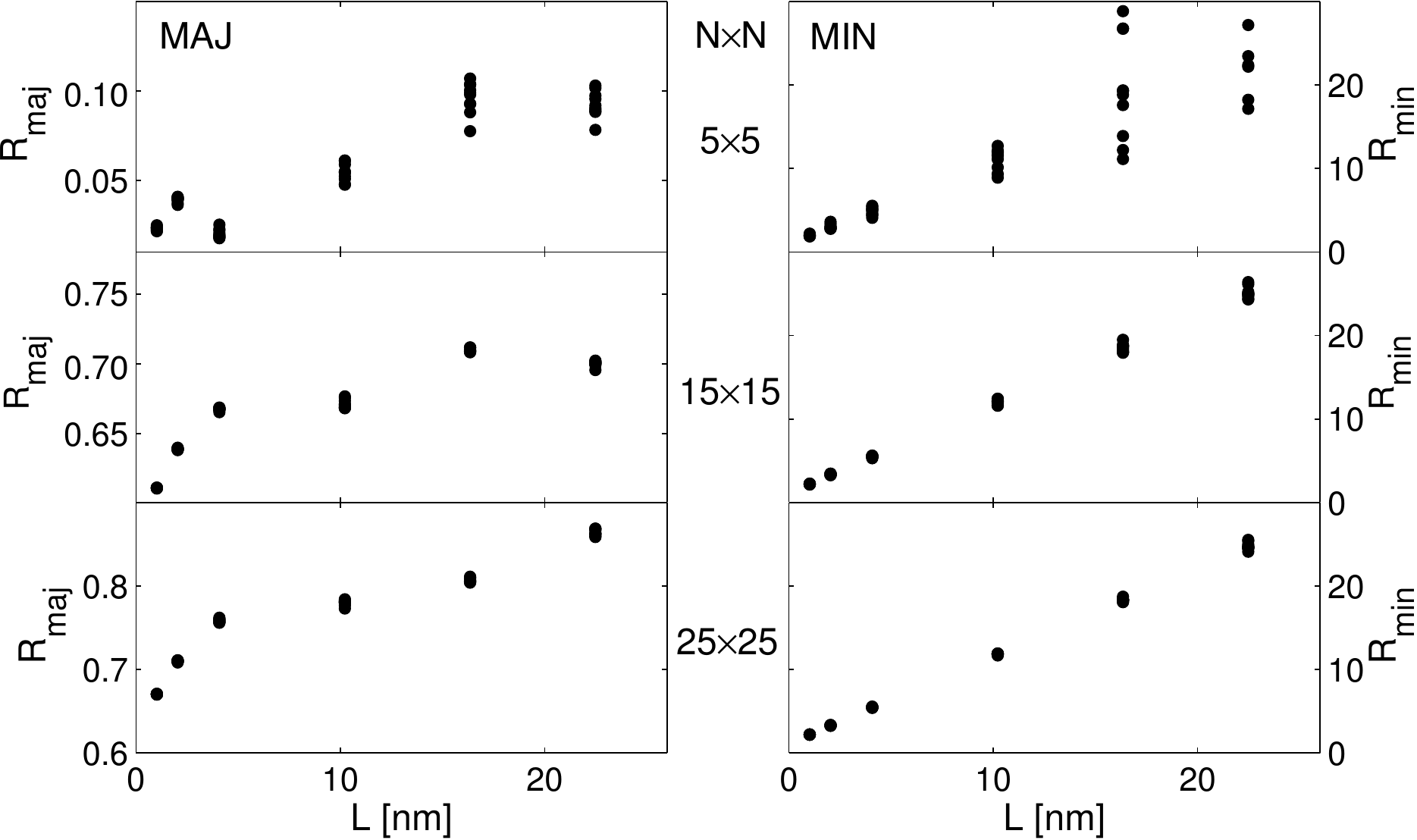}
\caption{Resistance in f$\Omega\,$m$^2$ of a $\rm Cu|Py|Cu$ structure as a function of the thickness $L$ of Py for $\Gamma$-point calculations for $5\times 5$ (top), $15\times 15$ (middle), and $25\times 25$ (bottom) lateral supercells, for majority (left column) and minority (right column) spins, with 10 random configurations of chemical disorder per thickness. 
}
\label{fig:scgammares}
\end{figure}

By assuming lattice periodicity parallel to the interface, the wave functions can be characterized by a wave vector ${\bf k}_{\parallel}$ in a 2D BZ no matter how large the period might be. In the thermodynamic limit, this BZ becomes vanishingly small, the band dispersion becomes negligible and neither BZ sampling nor configuration averaging over configurations of disorder should be necessary. We explore this limit in Fig.~\ref{fig:scgammares} where the dependence of the resistance of a $\rm Cu|Py|Cu$ system is shown as a function of the Py slab thickness $L$ for $5\times5$, $15\times15$ and $25\times25$ supercells without SOC, i.e. examining the majority and minority spin subsystems separately and neglecting the effects of periodicity entirely by using only ${\bf k}_{\parallel}=(0,0)\equiv \Gamma$.

\begin{figure}[b]
\begin{center}
\includegraphics[width=0.9\columnwidth]{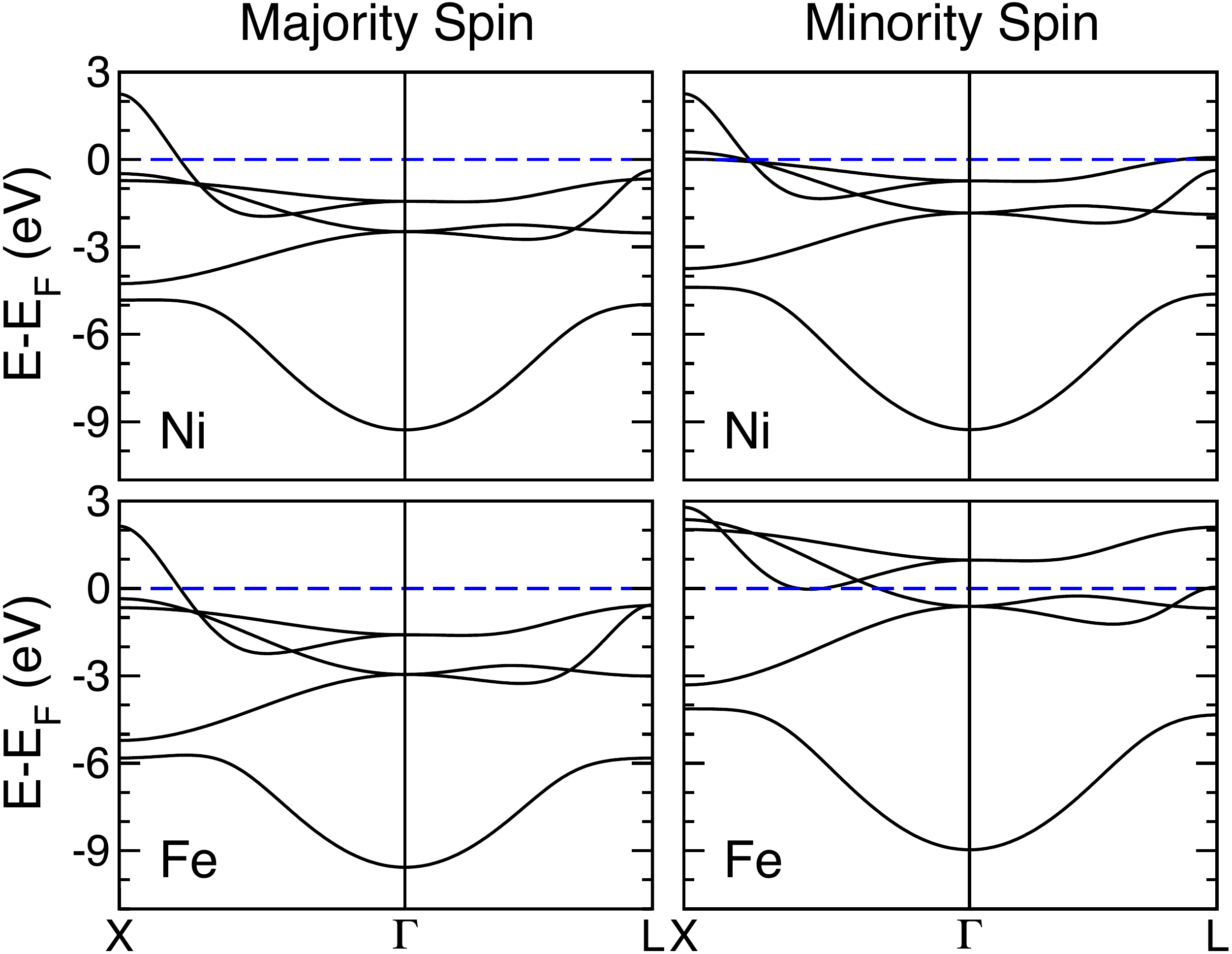}
\caption{Band structures calculated with the Ni and Fe AS potentials and Fermi energy that were calculated self-consistently for Ni$_{80}$Fe$_{20}$ using the coherent potential approximation. The same AS radii were used for Ni and Fe.}
\label{fig:py}
\vspace{-1em}
\end{center}
\vspace{-1em}
\end{figure}

The first feature we observe in Fig.~\ref{fig:scgammares} is a roughly two-order-of-magnitude difference between the resistances of majority and minority spins. Such a difference is qualitatively consistent with what is known about the mean-free paths of electrons in the two spin-channels \cite{Dieny:jmmm94}. For minority spins, this is extremely short with reported values in the range $4-8$~{\AA} while for majority spins it is much larger, in the range $50-200$~{\AA} \cite{Dieny:epl92, Gurney:prl93}. We can understand this  \cite{YuanZ:prl14} in terms of the energy bands that were calculated for fcc Fe and Ni using the AS potentials calculated self-consistently for permalloy with the CPA \cite{Soven:pr67, Turek:97}, shown in Fig.~\ref{fig:py}. At the Fermi energy, the majority-spin bands for Ni and Fe are almost identical so that in a disordered alloy the majority-spin electrons see essentially the same potentials on all lattice sites and are only very weakly scattered by the randomly distributed Ni and Fe potentials. In contrast, the minority-spin bands are quite different for Ni and Fe which can be understood in terms of the different exchange splittings; the magnetic moments calculated for Ni and Fe in permalloy in the CPA are 0.63 and 2.61$\,\mu_B$, respectively. The random distribution of Ni and Fe potentials in permalloy then leads to strong scattering of minority-spin electrons in transport. This picture is consistent with previous calculations of the resistivity and Bloch spectral function of permalloy \cite{Banhart:prb97, Khmelevskyi:prb03, Turek:prb12}.

As we approach the diffusive limit, we expect to see a linear dependence of the  resistance on the slab thickness. The minority spins clearly exhibit this  behaviour even for the smallest supercell size studied, $N \times N=5 \times 5$. Increasing $N$ only reduces the spread between results for different configurations of alloy disorder. For the majority spins the situation is rather different: the resistance depends non-linearly on $L$ with notable oscillations that we attribute to constructive and destructive interference of electron waves in the Fabry-Perot-like Cu$|$Py$|$Cu cavity. Only for thick Py or large supercells do the oscillations vanish, restoring the expected linear dependence of the resistance for $L>10\,$nm. 

In the case of a $5\times 5$ supercell, with only 5 layers of $\rm Py$ ($\sim 1\,$nm, the smallest Py slab thickness $L$ in Fig.~\ref{fig:scgammares}), the system size is already comparable to the mean free path of electrons in the minority spin-channel, and we are in the diffusive limit. For the majority-spin channel, the lateral dimensions of the slab only begin to approach the reported mean free path \cite{Dieny:epl92, Gurney:prl93} for a $25\times 25$ supercell and even for such large supercells the resistance only shows diffusive (linear) behaviour when the length of the slab is larger than the mean free path.

\begin{figure}[b]
\centering
 \includegraphics[width=0.45\textwidth]{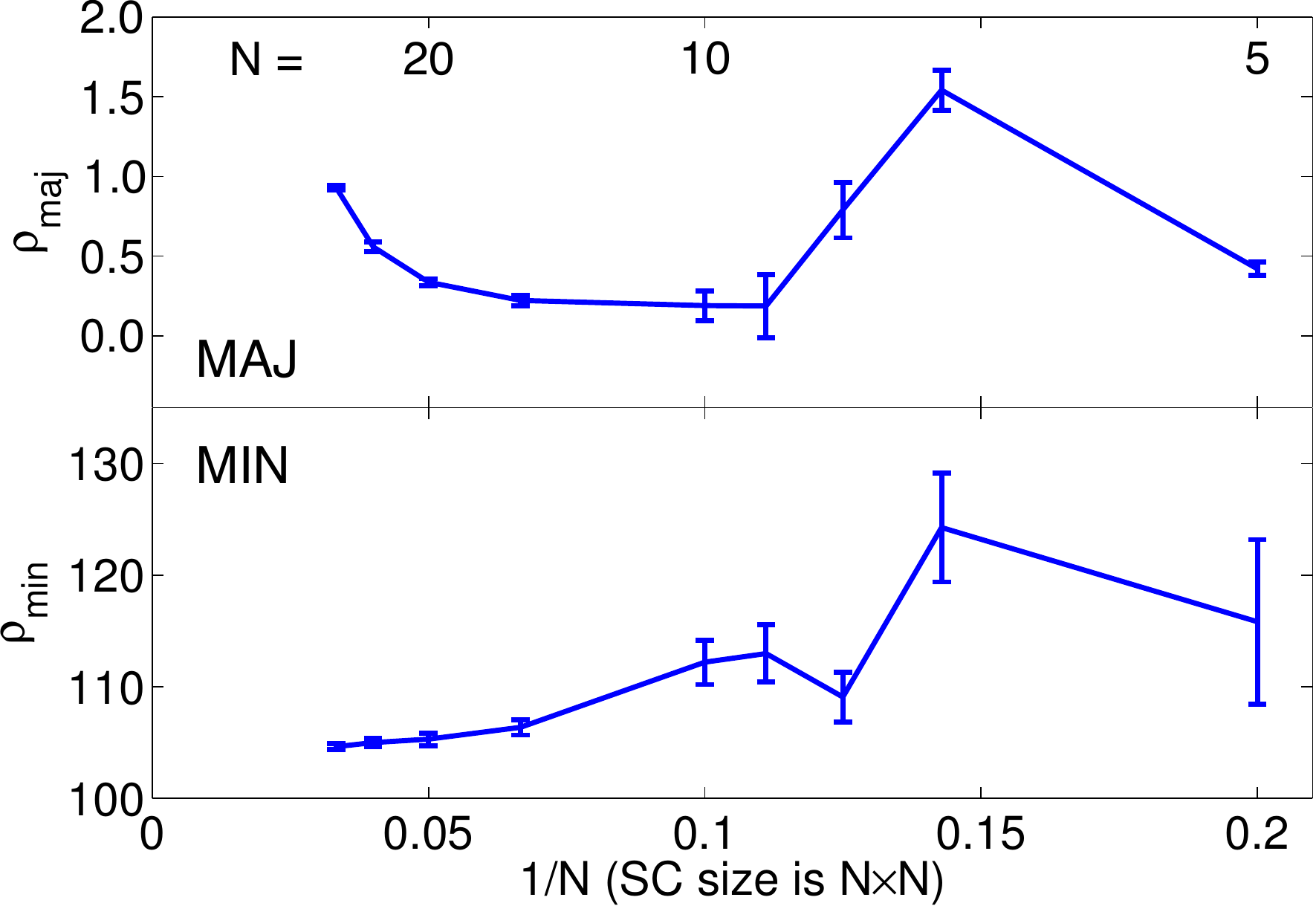}
\caption{Resistivity in $\mu\Omega\,\mathrm{cm}$ of $\rm Py$ as a function of the supercell size for $\Gamma$-point calculations: for majority (upper panel) and minority (lower panel) spins. The ``error bars'' measure the spread on averaging different configurations of alloy disorder. 
}
\label{fig:gamresist}
\end{figure}

Once the system is large enough to approach the diffusive limit and the length dependence of the resistance becomes linear, a resistivity $\rho^{\Gamma}$ can be determined from the slope of $R$ versus $L$. The dependence of $\rho^{\Gamma}$ on the supercell size $N$ is shown in Fig.~\ref{fig:gamresist}. One can see that in the minority spin case the resistivity is converged with a negligible error bar to $\rho_{\rm min}^{\Gamma}\sim105\,\mu\Omega\,$cm when the linear dimensions of the supercell are much larger than the mean free path. For the majority spin case, as discussed above, we are not yet in the diffusive limit and quantum (interference) effects are still observable. For sufficiently large values of $L$ we extract resistivity values of order $\rho_{\rm maj}^{\Gamma}\sim 0.5\,\mu\Omega\,$cm. Though the ``error bar'' that results from configuration averaging is small, there is still a strong dependence on the size of the lateral supercell. 

\subsubsection{Small supercell with integration over 2D BZ}
\label{sec:resBZ}

Although $\Gamma$-point calculations with a large supercell might be the most direct way of simulating a diffusive medium, the computational cost is very high. We therefore study the effect of improving the sampling of the wave functions by making use of Bloch's theorem. For an $N \times N$ lateral supercell, we calculate the transmission using a $Q \times Q $ set of ${\bf k}_{\parallel}$  points in the 2D BZ associated with the supercell. The total transmission is given by summation of partial transmissions. The same effect could be obtained for a $QN \times QN$ supercell with $\Gamma$ point sampling only (and with disorder in an $N \times N$ unit cell artificially repeated $Q \times Q$ times). 

\begin{figure}[b]
\centering
 \includegraphics[width=0.47\textwidth]{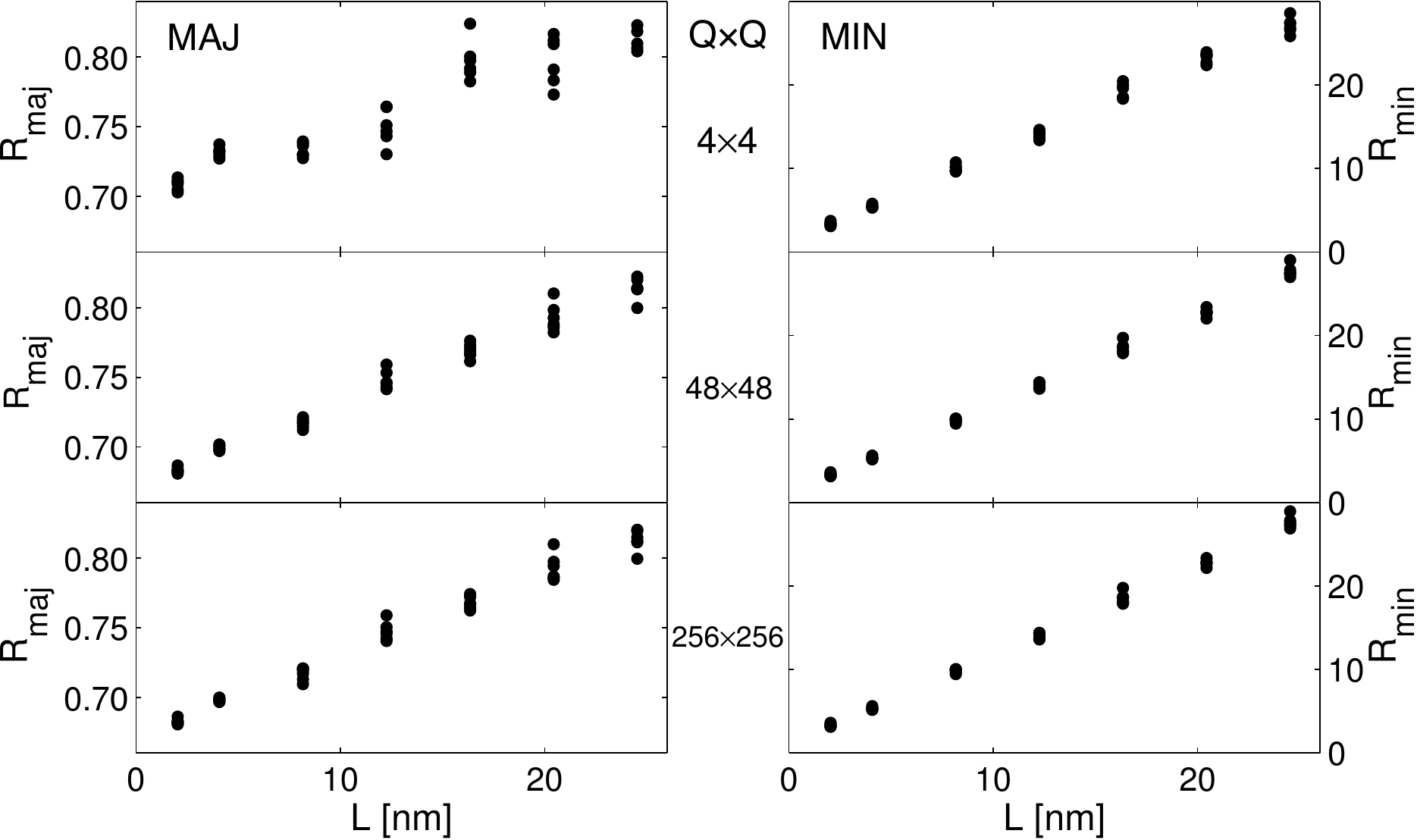}
\caption{Resistance in f$\Omega\,\mathrm m^2$of a Py slab in a $\rm Cu|Py|Cu$ scattering geometry as a function of the slab thickness $L$ for a $5\times5$ supercell with different k-point samplings of the Brillouin zone. The normalised area of the 2D element used in the BZ summation is defined as $\Delta^2 {\bf k}_{||}=A_{BZ}/Q^2$ where $A_{BZ}$ is the area of the downfolded supercell BZ. Results are shown for $Q=4$ (top), 48 (middle), and 256 (bottom) for majority (left column) and minority (right column) spins, with 6 random configurations of chemical disorder for every value of $L$. 
}
 \label{fig:bzressc5}
\end{figure}

The resistance of a Py slab is shown in Fig.~\ref{fig:bzressc5} for a $5\times 5$ supercell as a function of the thickness $L$ for different $Q \times Q$ samplings of the 2D BZ. As discussed in the previous section, for a $5\times 5$ supercell, especially for majority spins, we were far from the diffusive limit and the $\Gamma$ point picture was dominated by interference effects; these are still visible in the top left subplot in Fig.~\ref{fig:bzressc5} when only $4 \times 4$ k points are used for the BZ integration. Nevertheless, even though the resistance displays oscillatory behaviour for any individual k point, these oscillations average out with increasing BZ sampling density resulting in a substantially linear dependence visible in the bottom left subplot of Fig.~\ref{fig:bzressc5} for a sampling of $256 \times 256$ k-points. For minority spins the picture is simpler, as we already approach the diffusive limit for a $5\times 5$ supercell and therefore even a quite small BZ sampling results in a very linear dependence. Once the sampling is dense enough and we observe linear diffusive behaviour, we can extract values of the Py resistivity from the slope of $R(L)$. In the limit that $L \rightarrow 0$, the resistance does not vanish because of the interface resistance and the finite (Sharvin) conductance of the ideal leads.

\begin{figure}[!ht]
\centering
 \includegraphics[width=0.47\textwidth]{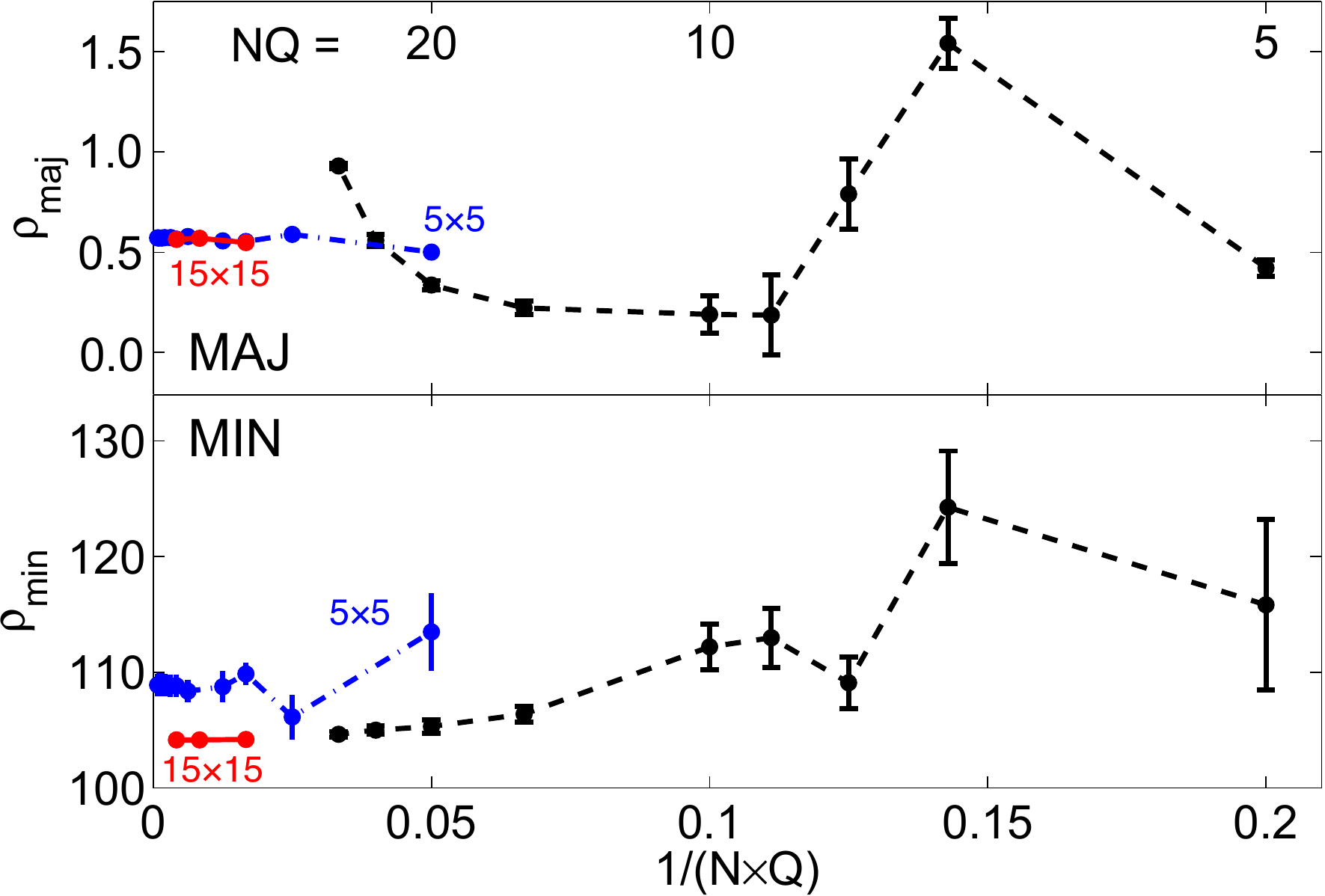}
\caption{Resistivity in $\mu\Omega\,\mathrm{cm}$ of Py as a function of equivalent k-point sampling ($N \times Q$) in which the downfolded 2D BZ with area $A_{BZ}$ for an $N \times N$  supercell is sampled with BZ element $\Delta^2 {\bf k}_{||}=A_{BZ}/Q^2$ for $5\times 5$ (dash-dotted blue line) and $15 \times 15$ (solid red line) supercells. For comparison the results of $\Gamma$-only calculations ($Q=1)$ with variable supercell size are also shown (dashed black line). Results for majority and minority spins are shown in the upper and lower panels, respectively. 
The converged values $\rho_{\rm maj}=0.57~\mu\Omega$~cm and $\rho_{\rm min}=105-109~\mu\Omega$~cm are in very good agreement with the calculated values in the literature \cite{Banhart:prb97} around 0.6 and 100 $\mu\Omega$~cm, respectively.
}
 \label{fig:allresist}
\end{figure}

To investigate the convergence of this procedure and compare with the $\Gamma$-only limit, we plot in Fig.~\ref{fig:allresist} the resistivity calculated for $5 \times 5$ and $15 \times 15$ supercells as a function of the equivalent BZ sampling $NQ \times NQ$ together with the results for the $\Gamma$-only calculations for an $N\times N$ supercell from Fig.~\ref{fig:gamresist}. For majority-spin electrons $\Gamma$-only calculations are dominated by interference effects and far from convergence so an acceptable estimate of the resistivity could not be made. When the BZ sampling is increased, then the results for both $5 \times 5$ and $15 \times 15$ supercells converge rapidly to the same value of $0.57 \pm 0.01 \, \mu\Omega\,$cm for majority spin electrons, suggesting that this is the true calculated (albeit experimentally unobservable because SOC has been omitted) resistivity. 
For minority spins, the BZ-integrated $15 \times 15$ supercell result provides us with a converged value of $105 \pm 1 \, \mu\Omega\,$cm, which is similar to the $\Gamma$-only result with a large supercell. The converged value for a $5 \times 5$ supercell is $4\%$ larger ($109 \pm 1 \, \mu\Omega\,$cm), which can be attributed to the error that results from the limited averaging of configuration space possible with a limited supercell size; for a mean-free-path of $1\,$nm, a volume containing only $5 \times 5 \times 5$ atoms is sampled by a minority spin electron before undergoing a collision. For our present purposes, this error is not big enough to justify the greater expense associated with larger supercells and we will limit ourselves to the $5 \times 5$ supercell with $32\times 32$ 2D BZ sampling throughout the rest of this paper. This is something which should be born in mind when comparing to experiment where calculations should be explicitly tested for convergence with respect to lateral supercell size as well as BZ sampling.
Additional tests of other aspects of the numerical implementation of the method can be found in Appendix~\ref{sec:extratests}.

\subsection{Resistivity calculations with SOC}
\label{res:resSOC}

\begin{figure}[t] 
  \centering
  \includegraphics[width=0.47\textwidth]{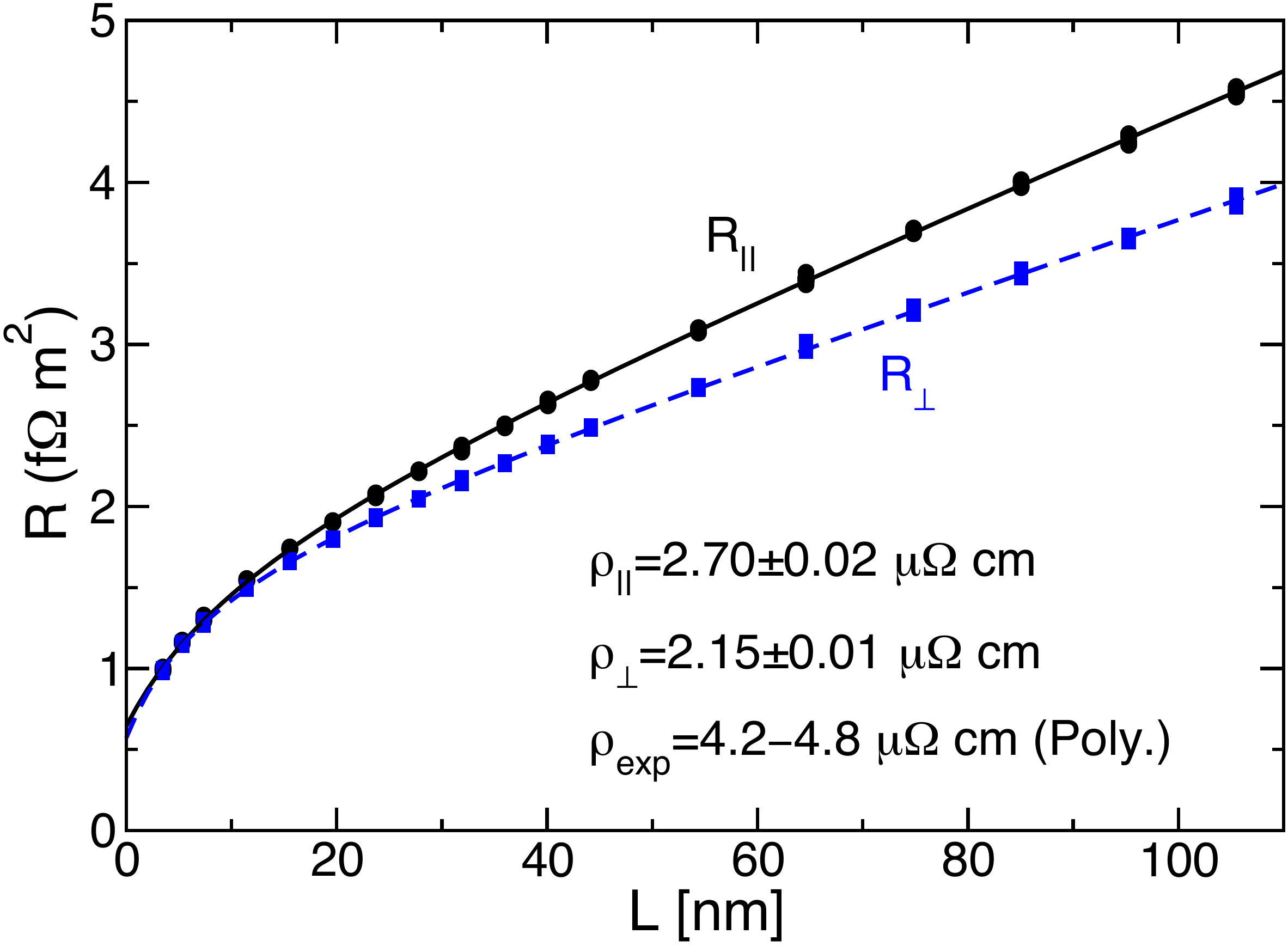}
  \caption{Resistance calculated for $\rm Cu|Py|Cu$ using a $5\times5$ supercell with SOC as a function of the layer thickness $L$ with the magnetization parallel to ($R_{\parallel}$: dots, solid line) and perpendicular to ($R_{\bot}$: squares, dashed line) the current direction. Calculated results are shown by symbols while the lines are fits using the 2CSR model.
The fitted resistivities are nearly a factor of two smaller than experimental values in the range 4.2--4.8 $\mu\Omega$ cm measured for polycrystalline samples at low temperature, \cite{Smit:phys51, McGuire:ieeem75, Jaoul:jmmm77, Cadeville:jpf73} where the presence of grain boundaries can increase the resistivity \cite{Banhart:epl95, Zhou:ssc10}.
  }
\label{fig:resistance}
\end{figure}

Fig.~\ref{fig:resistance} shows the thickness dependence of the Py layer resistance where SOC was included with the magnetization perpendicular to the current direction ($R_{\bot}$) and parallel to it ($R_{\parallel}$). What is most striking about these results compared to those without SOC in Fig.~\ref{fig:bzressc5} is the nonlinearity of $R(L)$. When a current of electrons is injected from the Cu lead on the left into disordered Py, they need not scatter immediately at the interface but do so on a length scale measured in terms of the elastic mean free path. However, we do not see any evidence for such an effect in the absence of SOC (Fig.~\ref{fig:bzressc5}) and  there is no good reason why SOC should greatly alter this. When we include SOC, spin is no longer a good quantum number and the unpolarized current must adapt to the finite polarization of Py. This it does asymptotically on a length scale given by $l_{\rm sf}^{\rm Py}$, the spin-flip diffusion length in Py which was calculated in Ref.~\onlinecite{Starikov:prl10} to be $\sim 5.5$~nm in good agreement with values determined experimentally \cite{Bass:jmmm99, Bass:jpcm07}. However, in Fig.~\ref{fig:resistance} $R$ appears to be varying nonlinearly on a length scale much larger than this value of $l_{\rm sf}^{\rm Py}$. To understand why, we must return to calculations for Py without SOC.

\begin{figure}[t]
\centering
 \includegraphics[width=0.47\textwidth]{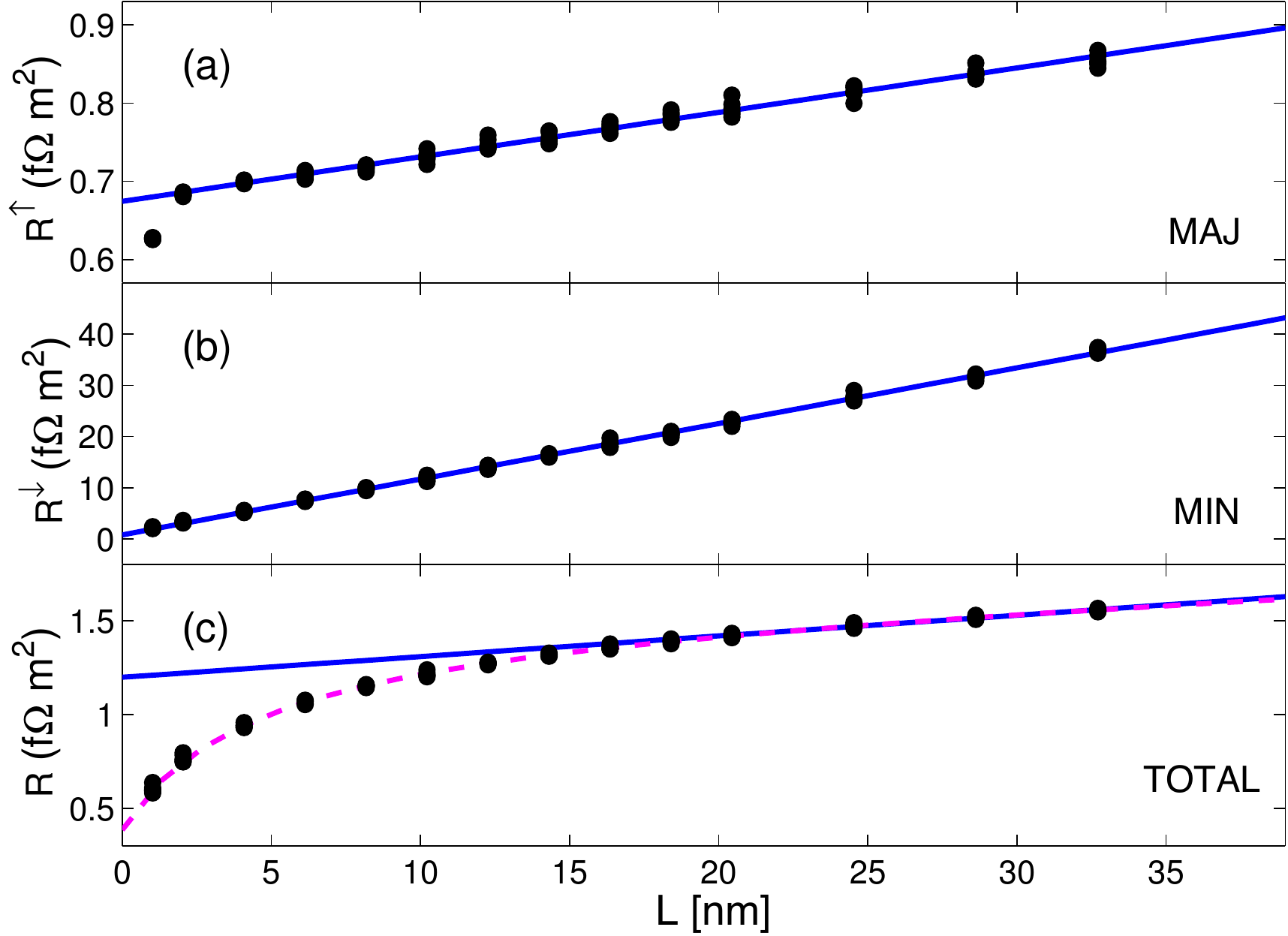}
\caption{Resistance calculated for $\rm Cu|Py|Cu$ for a $5\times5$ supercell without SOC as a function of the Py slab thickness  $L$  for (a) majority spin electrons, (b) minority spin electrons and (c) total. The multiple symbols for a given length are results for different configurations of disorder. Linear least-squares fits for $L>20 \,$nm are shown by solid lines, a non-linear least squares fit for total resistance is represented by a dashed line.}
 \label{fig:psres}
 \end{figure}

Within the 2CSR model the resistances of the individual spin channels are first determined and then added in parallel to determine the total. These individual spin resistances are shown in Fig.~\ref{fig:psres} for the same system as studied in Fig.~\ref{fig:resistance} but with the SOC switched off. 
The majority and minority spin resistances are perfectly linear (except for a small mean-free-path effect showing up for a Py thickness smaller than $2\,$nm in the majority spin channel). 
The total resistance shown in the bottom panel exhibits a curvature that is absent in the individual spin channels and in addition, the slope is about a factor of four smaller than with SOC included.
We can understand the curvature from \eqref{eq:FMresistance} or by \eqref{eq:FMres_simple}; we only expect to observe the linear behavior characteristic of Ohm's law when the third term on the right hand side of these equations is negligible compared to the other terms (or vanishes). This only happens when $\rho L \gg R_i$ (or $\beta=\gamma$).

It is instructive to try and extract a value for the resistivity from Fig.~\ref{fig:psres}(c) while pretending not to know the slopes of $\rho_{\rm maj}=0.57\pm0.01 \,\mu \Omega\,$cm and $\rho_{\rm min}=109 \,\mu \Omega\,$cm from 
Figs.~\ref{fig:psres}(a) and \ref{fig:psres}(b) that when combined result in $\rho_{\rm nonrel}=(\rho_{\rm maj}^{-1}+\rho_{\rm min}^{-1})^{-1}=0.567\pm0.009 \,\mu \Omega\,$cm. We can do this either by fitting the expression \eqref{eq:FMres_simple} to the calculated data or by studying systems so long that the contribution of the bulk resistivity to the total dominates the interface terms and can be extracted from the slope of $R(L)$. 
A linear fit of $R(L)$ for $L>20\,$nm yields a total resistivity of $1.08\pm0.07\,\mu \Omega\,$cm which is almost twice as large as the true value, $\rho_{\rm nonrel}$, demonstrating that the non-linear contribution from the interface terms in \eqref{eq:FMres_simple} is still not negligible even though $R(L)$ looks reasonably linear. Estimating the resistivity with higher accuracy using this approach requires calculations for much longer systems.

The alternative is to fit $R(L)$ using \eqref{eq:FMres_simple}. To ensure a stable fitting procedure we use an iteratively re-weighted least squares algorithm with bisquare weights\cite{Dumouchel:91}. In this case the estimated total resistivity is $0.53\pm0.05\,\mu \Omega\,$cm. Though this is in much better agreement with $\rho_{\rm nonrel}$, an additional error of 7\% has nevertheless been introduced and the errorbar itself has increased by a factor 5. 

Returning now to Fig.~\ref{fig:resistance}, we find that the $R(L)$ data with SOC shown as symbols can be fitted very well using \eqref{eq:FMres_simple} (solid and dashed lines) yielding resistivity values of $\rho_{\parallel}=2.70\pm0.02\, \mu\Omega\,$cm and $\rho_{\bot}=2.15\pm 0.01 \, \mu\Omega\,$cm. The average resistivity $\bar{\rho}=(\rho_{\parallel}+2 \rho_{\bot})/3=2.33\pm0.02 \, \mu\Omega\,$cm is a factor four larger \cite{Banhart:epl95, Banhart:prb97} than $\rho_{\rm nonrel}=0.567\pm0.009 \,\mu \Omega\,$cm. This compares reasonably well with CPA calculations performed within the Kubo-Greenwood formalism \cite{Banhart:epl95, Khmelevskyi:prb03, Turek:prb12} but is almost a factor of two smaller than  experimental values in the range $4.2 - 4.8 \, \mu\Omega\,$cm  measured for polycrystalline samples \cite{Smit:phys51, McGuire:ieeem75, Jaoul:jmmm77, Cadeville:jpf73}. 
Note that the resistivity can be significantly enhanced by the grain boundaries. \cite{Zhou:ssc10} The magnetoresistance anisotropy value we estimate is $(\rho_{\parallel}-\rho_{\bot})/\bar{\rho}\times100\%=24\pm1\%$ that compares reasonably with experimental values in the range $16-18\%$ \cite{Smit:phys51, McGuire:ieeem75, Jaoul:jmmm77} and previous theoretical estimates \cite{Banhart:epl95, Khmelevskyi:prb03, Turek:prb12}. 

Our calculations confirm the overall picture that in spite of its smallness for $3d$ materials, SOC plays an essential role in determining the transport properties of alloys when there is a very large difference between the resistivities of majority and minority spins in the absence of SOC. When temperature-induced lattice and spin disorder are included below, the bulk resistivity will increase and the curvature seen for low values of $L$ decrease; it will turn out that low temperature Py is peculiarly difficult to describe accurately in our real space approach because of the large mismatch between the two spin channels. 

\subsubsection*{Influence of the Interface}

Determining an alloy resistivity from calculations that include SOC using a 2CSR model that neglects it entirely is unsatisfactory. The 2CSR model does identify an important issue however, namely the essential role played by interfaces in the scattering formalism. Clearly the interface is obscuring the character of the bulk property we want to study by introducing a number of extraneous effects: mean-free-path effects at the interface, spin-dependent interface resistances, interface and bulk spin flipping required to bring the unpolarized current injected from the Cu lead into equilibrium with the spin-polarized current in Py. Since the asymptotic resistivity should be independent of the leads, ideally we would choose lead materials that are perfectly matched to the properties of the alloy we are studying with the same current polarization and minimal interface resistance. However, we are constrained in our choice of lead material to choose something with full lattice periodicity. We examine a number of possibilities in this section.

\begin{figure}[t]
\begin{center}
\includegraphics[width=0.9\columnwidth]{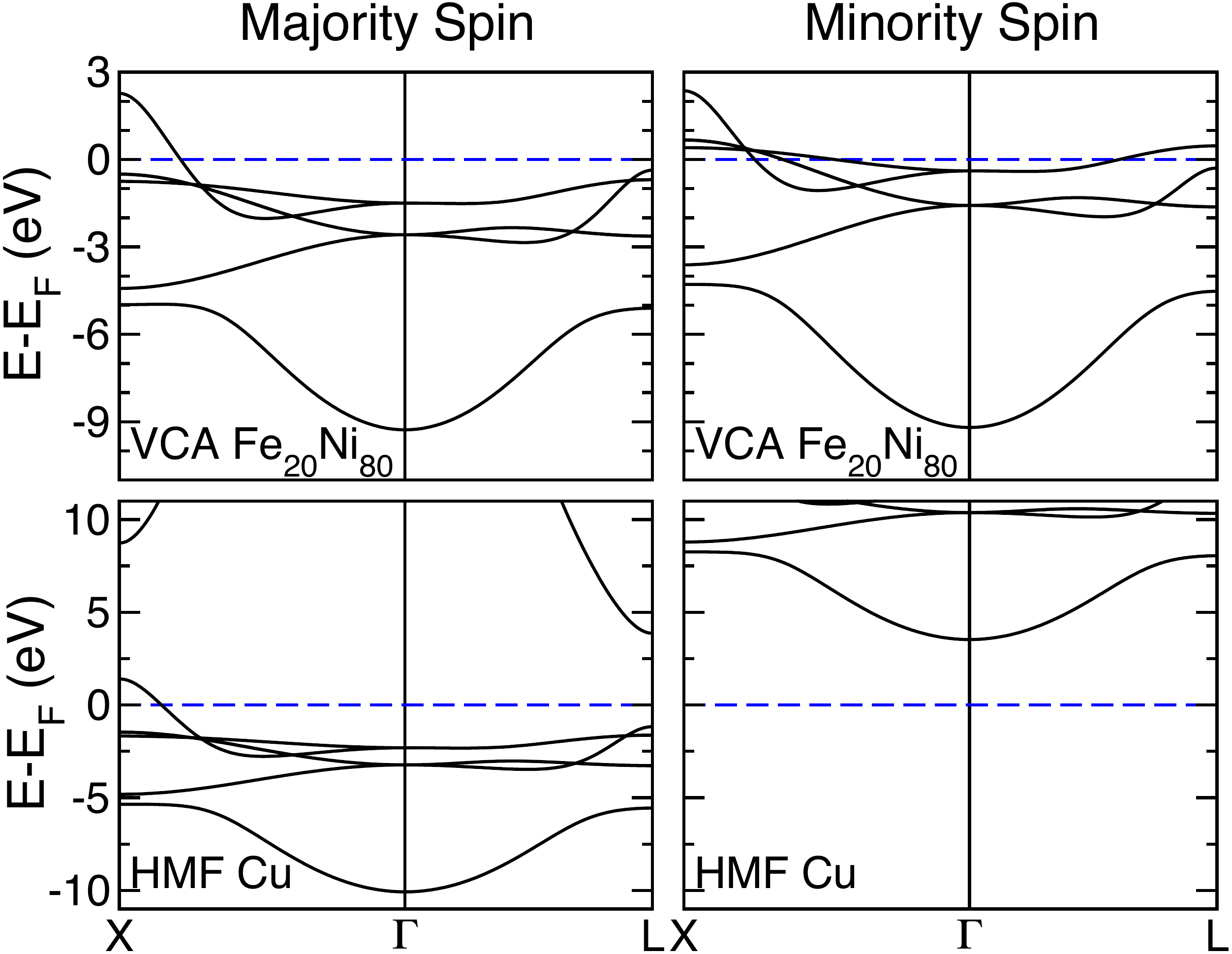}
\caption{Top row: self-consistent virtual crystal approximation band structure for a nuclear charge $Z_{\rm VCA}= (1-x) Z_{\rm Ni} + x Z_{\rm Fe}$. The majority spin states (lhs) are free electron like, the minority spin states have mainly $3d$ character (rhs).
Bottom row: majority (lhs) and minority (rhs) spin band structures of Cu in which a replusive constant potential of 1 Rydberg has been added to the minority spin potential, the effect being to remove all minority spin states from the Fermi energy.}
\label{fig:VCA}
\vspace{-1em}
\end{center}
\vspace{-1em}
\end{figure}

\begin{figure}[b]
\centering
 \includegraphics[width=0.47\textwidth]{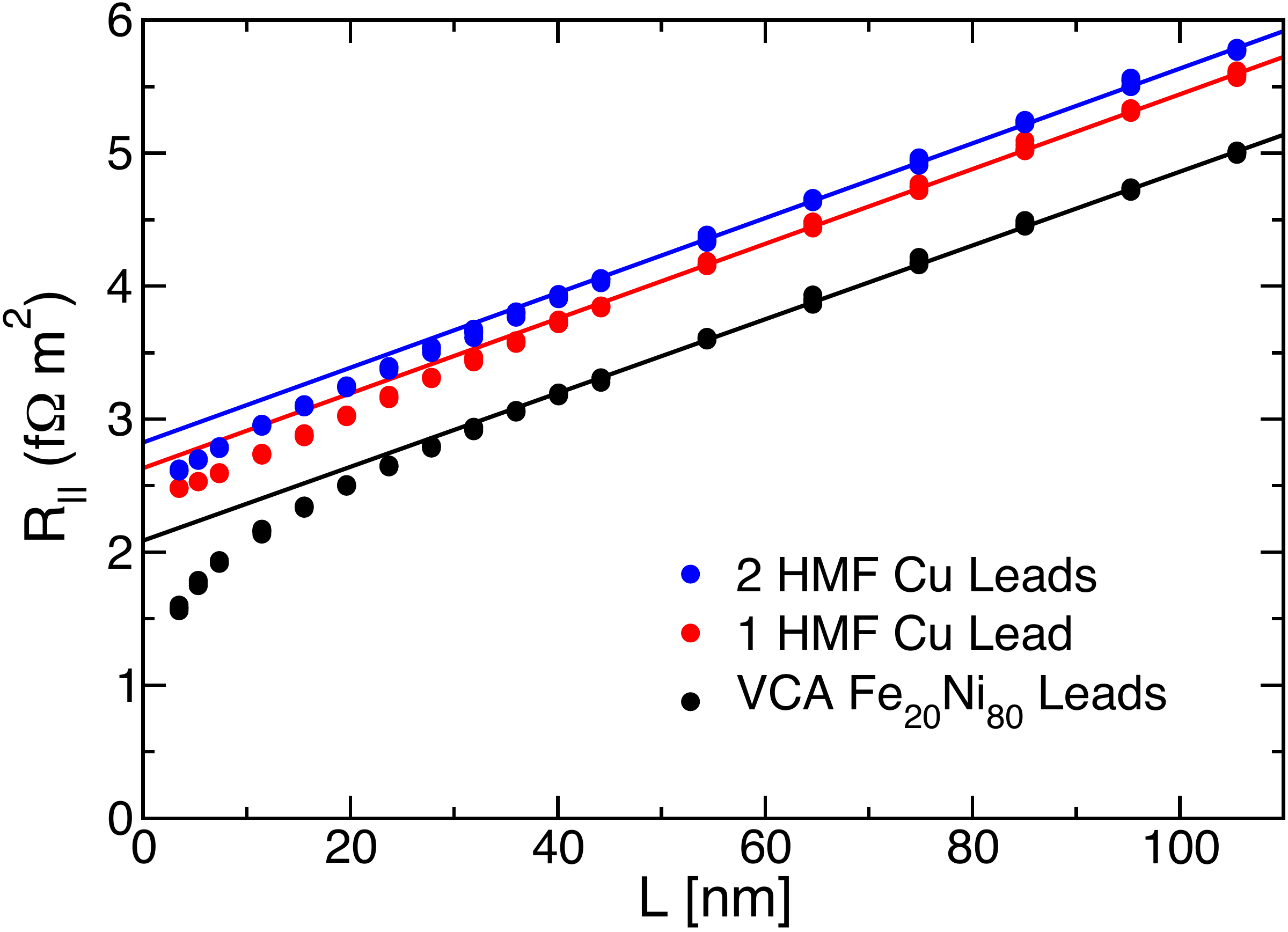}
\caption{Resistance calculated with SOC as a function of the Py slab thickness  $L$. The black dots are calculated using the VCA Ni$_{80}$Fe$_{20}$ leads, while the red (blue) dots are obtained using Cu leads with one (both) of the Cu leads replaced by the artificial HMF Cu. The solid lines are linear fits to the calculated values.}
 \label{fig:vcalead}
 \end{figure}

We could choose leads to be ordered alloys with the same chemical composition as Py. However, for an arbitrary Ni$_{1-x}$Fe$_x$ chemical composition this might require using impossibly large unit cells. Instead we adopt the virtual crystal approximation (VCA) \cite{Nordheim:adp31a, *Nordheim:adp31b} and construct artificial atoms with nuclear charge $Z_{\rm VCA}= (1-x) Z_{\rm Ni} + x Z_{\rm Fe}$ and a corresponding number of neutralizing valence electrons. A self-consistent calculation with this procedure for an fcc lattice with the same lattice constant as Py results in a magnetic moment of $1.06 \mu_B$; the corresponding bands are shown in Fig.~\ref{fig:VCA} (top row). We see that the majority spin states (lhs) are free electron like while the minority spin $d$ states (rhs) are partly occupied so we expect a better matching of the electronic structures at the interface that should result in a smaller interface resistance.
The result of calculating $R_{\parallel}(L)$ using these VCA leads is shown in Fig.~\ref{fig:vcalead} (black dots). The curvature for low values of $L$ is strongly reduced compared to Fig.~\ref{fig:resistance} and a resistivity value of $\rho_{\parallel}=2.76\pm 0.01 \, \mu\Omega\,$cm is directly extracted from the linear dependence. 

The procedure can be further refined by noting that the $1/L$ term in \eqref{eq:FMresistance} vanishes if $R_i^{\downarrow}\rightarrow \infty$ i.e., $\gamma=1$. This situation would correspond to using halfmetallic ferromagnetic (HMF) \cite{deGroot:prl83} leads. There is no need to actually use a ``real'' HMF, we can construct one from Cu leads by simply adding a strong ($\sim 1\,$Ry) repulsive potential to the minority spin Cu lead potentials to eliminate all minority spin states from the vicinity of the Fermi level (Fig.~\ref{fig:VCA}, lower panels). In this way only majority spins are injected into Py whose low temperature polarization we calculated to be $\beta\sim 0.9$ \cite{LiuY:prb15}. If we do this for the left-hand lead we obtain the results shown in Fig.~\ref{fig:vcalead} as red dots; constructing both leads in this way, we obtain the results shown as blue dots. In both cases, we achieve nearly ideal Ohmic behaviour. We attribute the small deviations from linearity for small values of $L$ to spin-flipping on a length scale of $l_{\rm sf}^{\rm Py}$ as the fully polarized injected electrons equilibrate to $\beta\sim 0.9$. The slopes are essentially identical within the error bars of the calculation, consistent with the  slope obtained with VCA leads with $\rho \sim 2.8 \, \mu \Omega\,$cm and only slightly larger than the value we extracted with unpolarized Cu leads, $\rho \sim 2.7 \, \mu \Omega\,$cm.  

All theoretical estimates of the permalloy resistivity are below the reported experimental range $\bar{\rho}=4.2 - 4.8 \mu\Omega\,$cm  \cite{Smit:phys51, McGuire:ieeem75, Cadeville:jpf73, Jaoul:jmmm77}. The discrepancy has been explained by noting that theoretical calculations have been carried out for monocrystalline, mono-domain permalloy, while experimental observations are made on polycrystalline samples \cite{Banhart:epl95}. In the latter case, additional scattering on the grain boundaries increases the resistivity. We also have no information about the domain structure of the experimental samples, but one can expect that for multi-domain samples the resistivity should further increase due to the additional scattering involved. 

\subsection{Gilbert damping}

\begin{figure}[t]
  \centering
  \includegraphics[width=0.47\textwidth]{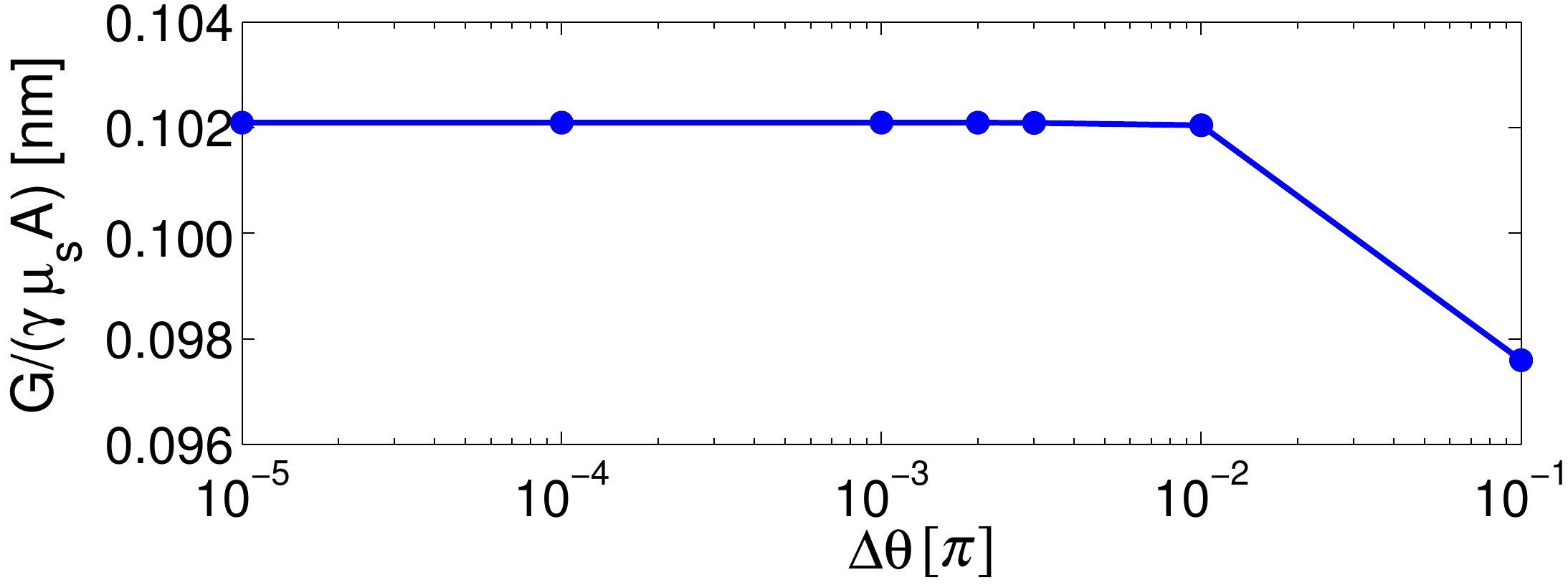}
  \caption{Gilbert damping of a 11.2\,nm thick layer of $\rm Ni_{80}Fe_{20}$ alloy as a function of deviation of the polar angle $\Delta\theta$ used for numerical differentiation. Calculations are performed using a $5\times5$ supercell. 
}
\label{fig:angconv}
\end{figure}

The Gilbert damping can be calculated by numerically differentiating the scattering matrices with respect to the magnetization orientation as formulated in Sec.~\ref{sec:gd}. The value of $\widetilde G$ resulting from the differentiation may depend on the choice of the finite but small value of $\Delta\theta$ chosen for the numerical procedure. An example is shown in Fig.~\ref{fig:angconv} for Py in the $\rm Cu|Py|Cu$ system. One can see that numerically $\widetilde G$ is very stable and does not depend on the choice of $\Delta\theta$ for variation of the polar angle over a large range,  indicating linear dependence of the scattering matrix on small variations of the magnetisation orientation.

\begin{figure}[b]
  \centering
  \includegraphics[width=0.45\textwidth]{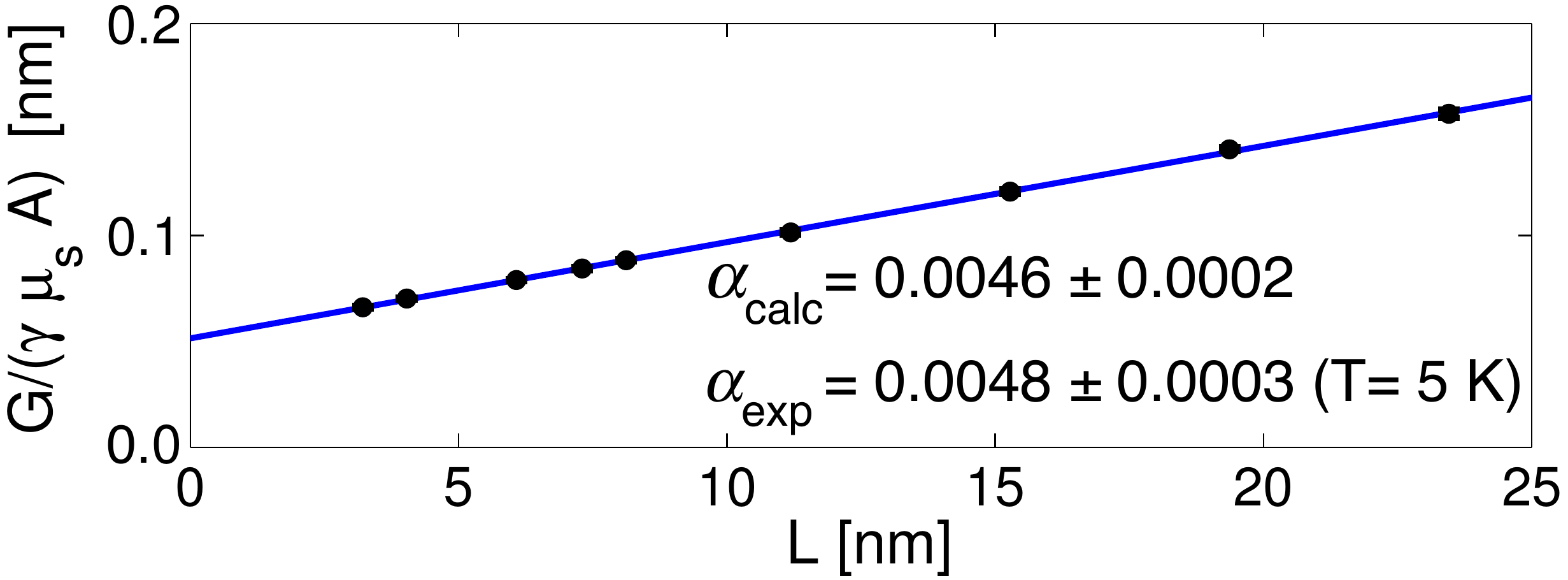}  
  \caption{Total damping of the $\rm Py$ slab as a function of its thickness $L$ \cite{footnote1}. The error bars, which measure the spread over different configurations of disorder for every thickness, are less than the marker size in this plot and thus nearly invisible. The damping extracted via a linear fitting (blue solid line), $\alpha_{\rm calc}=0.0046\pm0.0002$, is in excellent agreement with the experimental value reported at low temperature. \cite{Zhao:scr16}
  }
\label{fig:damppy}
\end{figure}

When the thickness of the Py layer is increased, the total damping of the system grows proportionally, as anticipated in \eqref{eq:dampfit} and demonstrated in  Fig.~\ref{fig:damppy}. Moreover, the strict linearity and negligible variation for different configurations of disorder indicate that the Gilbert damping is very insensitive to details of how the random chemical disorder in the alloy is modelled.
The value of $\alpha$ extracted from the slope of $\widetilde G(L)/(\gamma \mu_{s} A)$ is $0.0046\pm 0.0002$, which falls at the lower end of a range of values, $0.004-0.013$, reported in the literature for measurements at room temperature  \cite{Patton:jap75,  Mizukami:jmmm01, Mizukami:jjap01, Bailey:ieeem01, Nibarger:apl03, Ingvarsson:apl04, Nakamura:jjap04, Rantschler:ieeem05, Bonin:jap05, Lagae:jmmm05, Inaba:ieeem06, Oogane:jjap06, Kobayashi:ieeem09, Shaw:prb09, Luo:prb14, Yin:prb15, Zhao:scr16, Schoen:prb17b}. 
Very recently, measurements were carried out as a function of temperature from room temperature (RT) to low temperatures, decomposing the damping into bulk and interface contributions\cite{Zhao:scr16}. These yield a value for the bulk damping of $0.0048 \pm 0.0003$ at 5~K in remarkably good agreement with the value calculated above (that has been confirmed by subsequent coherent potential approximation calculations \cite{Mankovsky:prb13, Turek:prb15} within the error bars of the calculations). 

At room temperature, Zhao et al. reported a value of $\alpha=0.0055 \pm 0.0003$ that is at the low end of the $0.004-0.013$ range measured previously. An even more recent RT study of the Ni$_{1-x}$Fe$_x$ alloys as a function of $x$ reported\cite{Schoen:prb17b} values of $\alpha$ in essentially perfect agreement over the full concentration range  with the values calculated by Starikov et al. for $T=0$ \cite{Starikov:prl10}. Schoen et al. attributed the better agreement they obtained with theory to corrections they made (i) for interface damping enhancement \cite{Mizukami:jmmm01, Mizukami:jjap01, Mizukami:jmmm02, Ingvarsson:prb02} and (ii) for radiative damping. For Py they reported a RT value of $\alpha=0.0050$. Taken together with the Zhao et al. result, $\alpha=0.0055 \pm 0.0003$, this would seem to indicate a minimal temperature dependence of the magnetization damping that is striking in view of a factor five increase in the resistivity of Py over the same temperature range. To elucidate this (and the apparent disagreement between the $T=0$ calculations of Starikov et al. \cite{Starikov:prl10} and the older room temperature experiments), we undertook a study of the effect of thermally induced lattice and spin disorder on the resistivity and damping that will be the subject of the next section.  

\begin{figure}[t]
  \centering
  \includegraphics[width=0.45\textwidth]{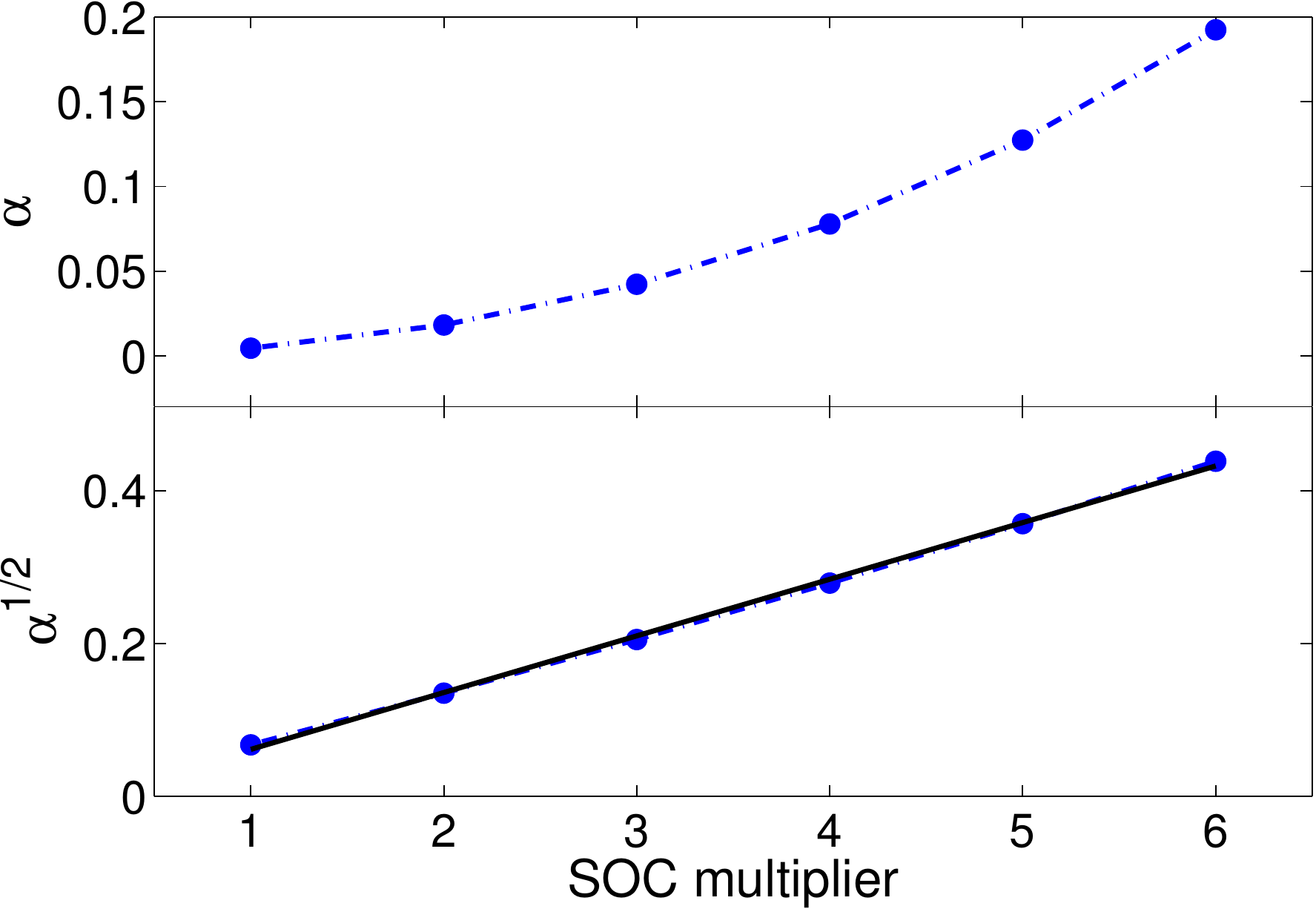}  
  \caption{Scaling of Gilbert damping with the SOC strength. The dots corresponds to the calculated results, and the solid line shows the quadratic fit. 
The quadratic dependence of $\alpha$ on the SOC strength is in agreement with a recent CPA calculation for Os-doped Py \cite{Mankovsky:prb13}.
}
\label{fig:soscaledamp}
\end{figure}

The ingredients contributing to the magnetization damping in the calculations are disorder and SOC; omitting either will lead to a vanishing bulk damping. In the next section we will examine what happens when we ``tune'' the amount of disorder. Before doing this, we investigate how the damping depends on the magnitude of the SOC term in \eqref{eq:hso} when we scale it with a parameter $\lambda$:  $H_{\rm so}\rightarrow \lambda H_{\rm so}$. From the results shown in Fig.~\ref{fig:soscaledamp} for Py it is clear that the  damping scales quadratically with the strength of the SOC.  This is the scaling expected for the strong interband scattering limit of the torque correlation model. Though only strictly applicable to ordered solids with well-defined band structures \cite{Kambersky:cjp76, Gilmore:prl07, Gilmore:jap08, Kambersky:prb07}, the strong interband scattering limit is the appropriate limit for a disordered alloy where momentum is not well defined.  

\subsection{Thermal lattice and spin disorder}
\label{sec:thermal}

To resolve the discrepancy between room temperature measurements \cite{Patton:jap75, Mizukami:jmmm01, Mizukami:jjap01, Bailey:ieeem01, Nibarger:apl03, Ingvarsson:apl04, Nakamura:jjap04, Rantschler:ieeem05, Bonin:jap05, Lagae:jmmm05, Inaba:ieeem06, Oogane:jjap06, Kobayashi:ieeem09, Shaw:prb09, Luo:prb14, Yin:prb15, Schoen:prb17b} and $T=0$ calculated values of $\alpha$ for the Ni$_{1-x}$Fe$_x$ alloy system \cite{Starikov:prl10} and because we are aware of only a single low temperature measurement \cite{Zhao:scr16}, we extend to Py the method introduced in Ref.~\onlinecite{LiuY:prb11} to study the temperature dependence of damping in Fe, Co and Ni. Finite temperatures lead to displacements of the atoms from their equilibrium positions (lattice disorder) and to rotations of atomic magnetic moments away from their equilibrium orientations. A correct theoretical description of temperature effects in solids would begin with the fundamental lattice and spin excitations (phonons and magnons, respectively) and the occupancy of these excitations \cite{LiuY:prb15} but this lies outside the scope of the present paper. Instead we apply a simpler scheme of uncorrelated Gaussian atomic and spin displacements \cite{LiuY:prb11} to study the effects of thermal lattice and spin disorder on the resistivity and damping of Py.

\begin{figure}[b]
  \centering
  \includegraphics[width=0.45\textwidth]{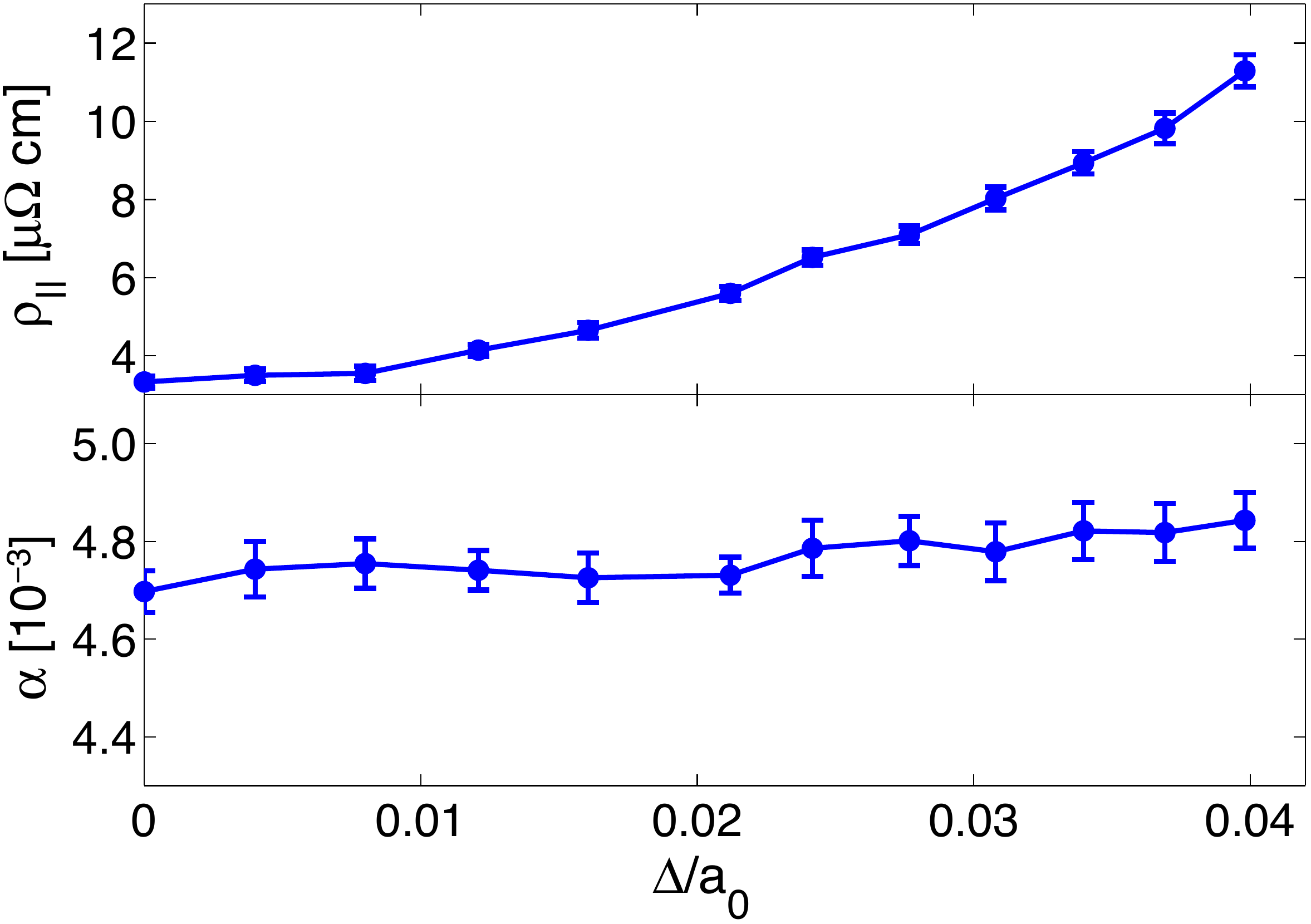}  
\caption{Resistivity $\rho_{\parallel}$ (upper panel) and damping parameter (lower panel) in $\rm Ni_{80}Fe_{20}$ alloy as a function of the rms displacment of atoms from their equilibrium positions (in units of the lattice constant $a_0$).} 
\label{fig:thermal}
\end{figure}

\begin{figure}[b] 
  \centering
  \includegraphics[width=0.47\textwidth]{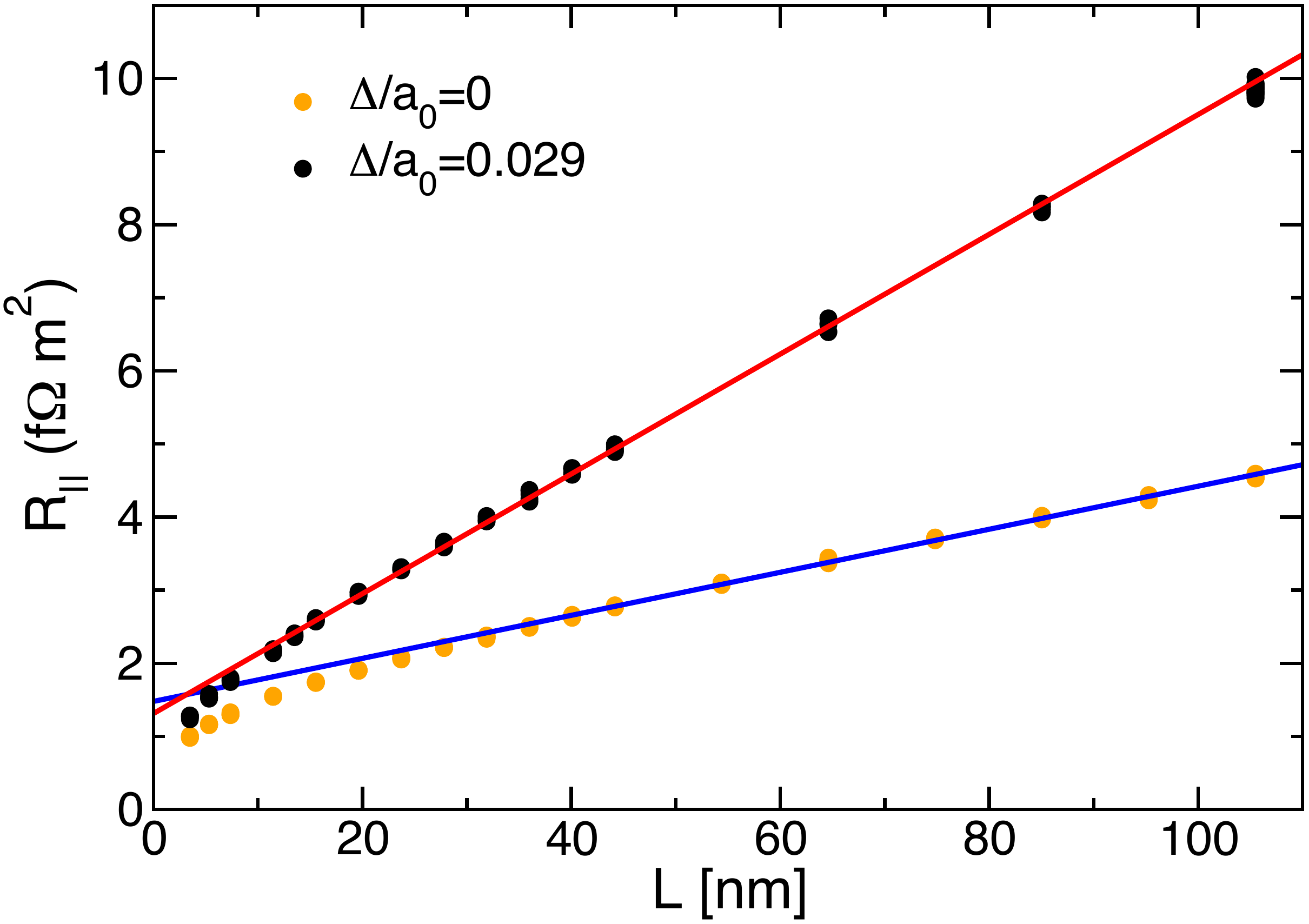}
  \caption{Resistance calculated for $\rm Cu|Py|Cu$ with lattice disorder corresponding to $\Delta/a_0=0.029$ in a $5\times5$ supercell and including SOC as a function of the layer thickness $L$. The magnetization is parallel to the current direction. The results without lattice disorder are included for comparison. The solid lines illustrate the linear fits to the calculated values.}
\label{fig:resistanceRT}
\end{figure}

We describe lattice disorder in terms of independent random displacements ${\bf u}_i$ of the $N_{\mathcal S}$ atoms in the scattering region labelled $i$ from their equilibrium positions ${\bf R}_i$ i.e., we describe the lattice as a collection of independent harmonic oscillators, see Fig.~\ref{fig:displacement}. The displacements ${\bf u}_i$ are distributed normally with rms displacement $\Delta=\sqrt{\sum_i u_i^2/N_{\mathcal S}}$. As shown in Fig.~\ref{fig:thermal}, increasing $\Delta$ leads to increased scattering and increased resistivity. For $\Delta/a_0=0.029$ corresponding to a resistivity of 8.2$\, \mu \Omega \,$cm, the resistance $R(L)$ of the Cu$|$Py$|$Cu system (unpolarized Cu leads) is shown as a function of $L$ in Fig.~\ref{fig:resistanceRT}. The increased bulk resistivity leads to a substantial decrease of the curvature observed in Fig.~\ref{fig:resistance} underlining the peculiar difficulties presented by low temperature Py.

While a rms displacement of 4\% (in units of the lattice parameter $a_0$) is enough to cause an almost four-fold increase in the resistivity, the damping increases by only $\sim 5 \%$. Fig.~\ref{fig:thermal} therefore indicates that while the resistivity might depend strongly on additional structural sources of scattering in Py such as dislocations, grain boundaries etc., the damping is expected to be insensitive to this additional disorder. The resistivity estimated theoretically for crystalline Py at zero temperature can be expected to be lower than the values determined experimentally for polycrystalline samples. 

The weak dependence of $\alpha$ on lattice disorder is consistent with the quadratic scaling of the damping with the SOC strength expected in the strong interband scattering limit of the torque correlation model. The electronic structure of Py strongly resembles that of clean Ni in sofar as the majority spin $d$ band is filled (seen clearly in the VCA, Fig.~\ref{fig:VCA}); it is known from experiment\cite{Bhagat:prb74} and TCM calculations in the strong interband scattering limit\cite{Kambersky:cjp76, Gilmore:prl07, Kambersky:prb07} that the damping parameter of Ni depends only weakly on the relaxation time in this limit.

\begin{figure}[t]
  \centering
  \includegraphics[width=0.45\textwidth]{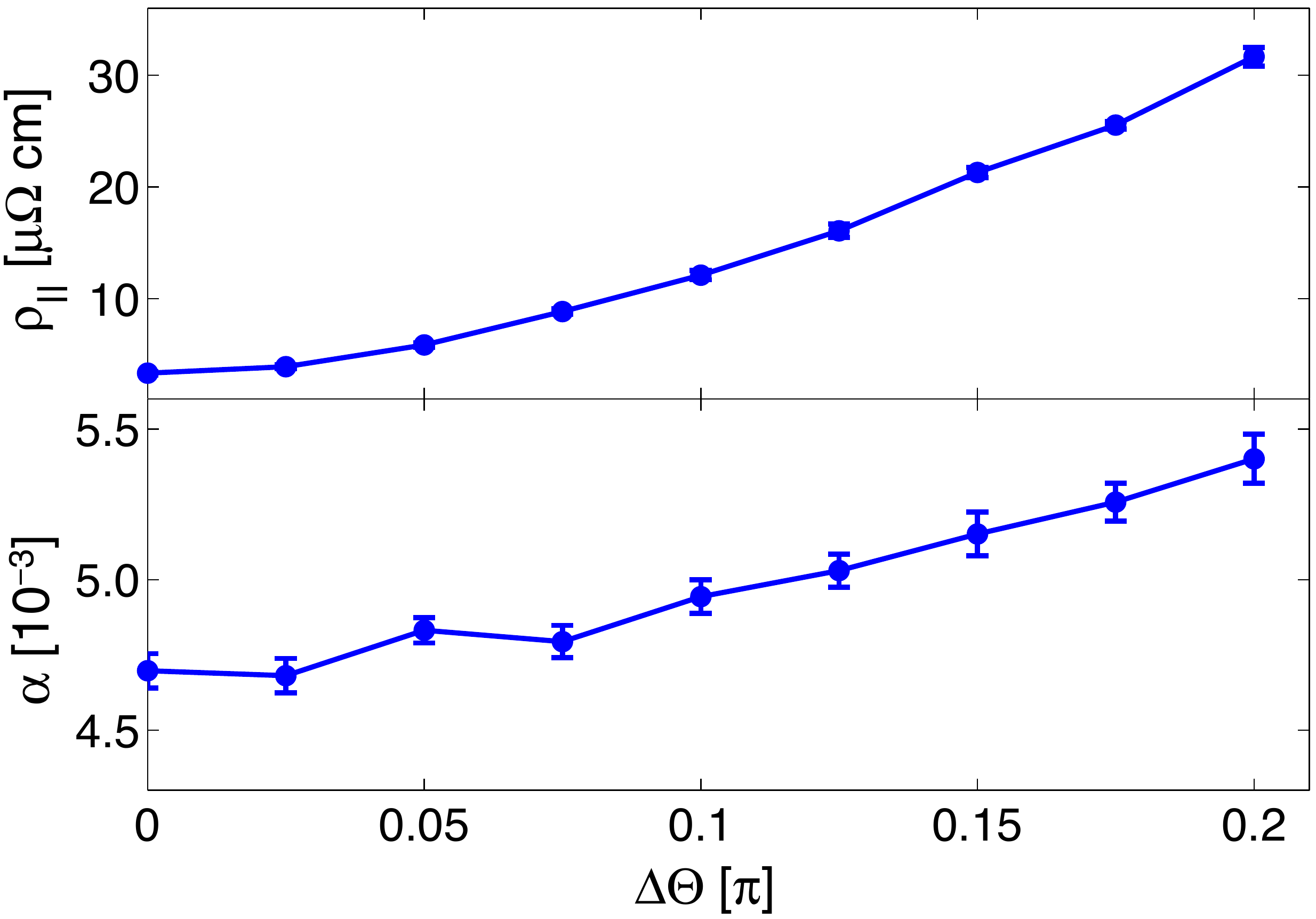}  
\caption{Resistivity (upper) and damping parameter (lower) in Ni$_{80}$Fe$_{20}$ alloy as a function of rms deviation of angle $\theta_i$ between atomic magnetic moments and macrospin. 
}
\label{fig:magnon}
\end{figure}

Reverting to the ideal crystal structure, we next introduce a certain level of spin disorder, as sketched in Fig.~\ref{fig:spindis}, by tilting the atomic moments randomly from their equilibrium orientations through angles $\theta_i$ that are assumed to be distributed normally with a rms tilt angle $\Delta\Theta=\sqrt{\sum_i\theta_i^2/N_{\mathcal S}}$. The results plotted in Fig.~\ref{fig:magnon} show that the resistivity depends strongly on spin disorder (suggesting that measured resistivity values should also depend on the domain structure of the samples); for the range of $\Delta\Theta$ shown, it increases by a factor of almost ten. Compared to the lattice disorder case, the damping parameter also increases more strongly, by almost 20\%. A major factor in this increase is, however, the reduced effective magnetisation density $\mu_s$ in \eqref{eq:dampfit} as $\Delta\Theta$ increases. Note that these calculations assume that the external magnetic field is negligibly weak. 

The qualitative picture sketched above can be improved by introducing quantitative measures for the rms displacements and rotations in terms of the temperature $T$. In the Debye model\cite{Willis:1975} the mean-square displacement of the $i$-th atom from equilibrium $\langle u^2_i \rangle$ depends on $T$ as
\begin{equation}
\langle u^2_i(T)\rangle=\Delta^2=\frac{3 \hbar^2 T}{m k \Theta^2_D}\left[\Phi\Big(\frac{\Theta_D}{T}\Big)+\frac{\Theta_D}{4T} \right]\:,
\label{eq:msd1}
\end{equation}
where $\Theta_D$ is the Debye temperature and $\Phi(x)$ is the Debye integral function
\begin{equation}
\Phi(x)=\frac{1}{x}\int_0^x\frac{y}{e^y -1}dy\:.
\label{eq:Debye}
\end{equation}
Expanding the integrand in \eqref{eq:Debye} as a power series in $y$ leads to $\Phi(x)=1-\frac{x}{4}+\frac{x^2}{36}+...$ so that in the high temperature limit where $x < 1$, $\Phi(x)\simeq 1-x/4$ and \eqref{eq:msd1} reduces to the classical statistics \cite{Willis:1975}
\begin{equation}
\Delta^2= \frac{3 \hbar^2 T}{m k \Theta^2_D}.
\label{eq:msd2}
\end{equation} 
In the low temperature limit $T \ll \Theta_D (x\gg 1)$, zero point motion (zpm) is dominant and $\Delta^2= \frac{3 \hbar^2}{4 m k \Theta_D}$. Because it does not contribute to give a low temperature resistivity, we neglect the zpm at $T=0$ but keep the complete Debye integral \eqref{eq:Debye} for finite temperature calculation.

To map spin disorder onto temperature, we can use a cubic spline to interpolate the experimental magnetisation \cite{Wakelin:ppsb53, Weber:jpcs63, *Weber:pr65} at an arbitrary temperature \cite{LiuY:prb15} or fit the experimental data using some empirical analytical function \cite{Kuzmin:prl05}. We assume that the magnetisation at finite temperature can be defined in terms of the mean value of the cosine of the tilting angle~$\theta$
\begin{equation}
M(T)=M_0 \langle \cos{\theta}\rangle\:.
\end{equation}
For atomic tilting angles $\theta_i$ distributed according to the Fisher distribution \cite{Fisher:prsa53} with probability density
\begin{equation}
P(\theta,\kappa)=\frac{\kappa \sin{\theta}\, e^{\kappa\cos{\theta}}}{2 \sinh{\kappa}}\;,\label{eq:fisher}
\end{equation}
the expectation value of the $z$-projection of the local magnetisation can be expressed via the distribution parameter $\kappa$ as $\langle\cos{\theta}\rangle=\coth{\kappa}-1/\kappa$ which is just the Brillouin function for large $J$  and $\kappa \sim 1/k_B T$. This results in a transcendental equation which relates temperature and magnetic moment distribution
\begin{equation}
\frac{M(T)}{M(0)}=\coth{\kappa}-\frac{1}{\kappa}\;.
\end{equation}
The azimuthal angle $\phi$ in \eqref{eq:rotop} defining the orientation of the projection of the magnetic moment in the $xy$ plane is assumed to be distributed uniformly in $[0,2\pi]$ which is reasonable for an isotropic material like Py.

The thermally induced random field satisfies the fluctuation dissipation theorem, i.e. the time-averaged correlation function of the random field is proportional to the Gilbert damping and depends on the temperature \cite{Foros:prb08, Foros:prb09, Xiao:prb10}. In our modeling of the spin disorder, we do not explicitly involve the random field but generate several snapshots of  magnetic configurations which are essentially ``independent'' of one another. So the implementation using the random distribution of the atomic magnetic moments \eqref{eq:fisher} does not violate the fluctuation dissipation theorem.

\begin{figure}[t]
  \centering
  \includegraphics[width=0.47\textwidth]{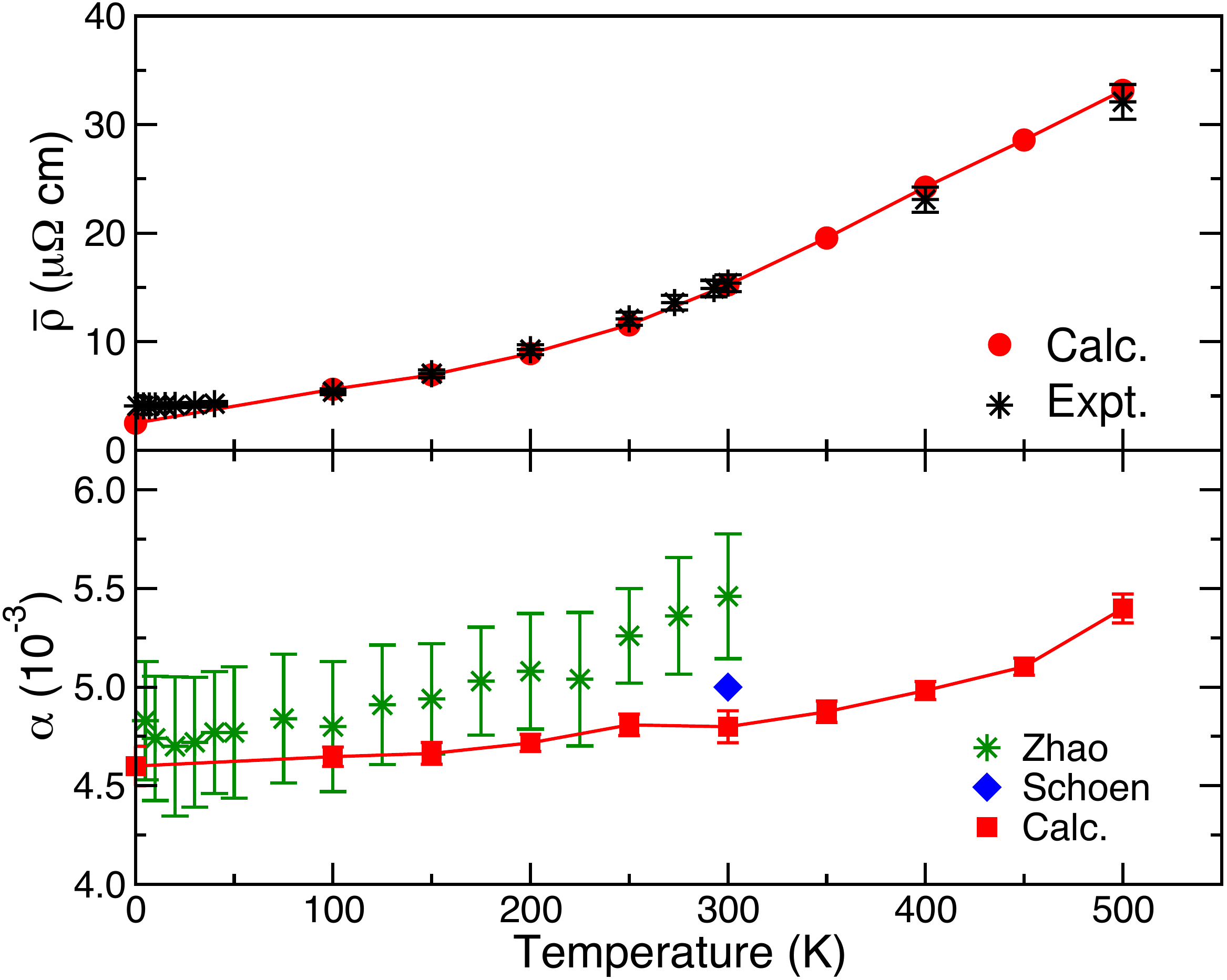}  
\caption{Resistivity (upper) and bulk damping (lower) of $\rm Ni_{80}Fe_{20}$ as a function of the temperature. The red symbols are the calculated results (the red line is a guide to the eye). The asterisks (black) in the resistivity plot correspond to tabulated average experimental values from \cite{Ho:jpcrd83}. The asterisks (green) in the damping plot are the data of Zhao et al.\cite{Zhao:scr16} 
The blue diamond is a room temperature measurement corrected for spin pumping and radiative contributions \cite{Schoen:prb17b}.
}
\label{fig:phonmag}
\end{figure}

The temperature dependent resistivity and damping calculated for Py using $\Theta_D=450K$ and the experimental magnetization \cite{Weber:jpcs63, *Weber:pr65} are compared with experiment in Fig.~\ref{fig:phonmag}. For temperatures around room-temperature, the model of independent harmonic oscillators describes phonon motion reasonably well and the phonon spectrum of transition metals is quite similar to the spectrum described by the Debye model. The agreement between the calculated and experimentally observed values of $\rho(T)$ \cite{Ho:jpcrd83} and of $\alpha(T)$ \cite{Zhao:scr16} is remarkably good. The results for the damping confirm an earlier report of an at best weak temperature dependence of $\alpha$ \cite{Counil:ieeem06}. A recent room temperature measurement \cite{Schoen:prb17b} of the damping containing corrections for spin pumping and radiative effects is in almost perfect agreement with our RT result. We note that the measurements of Zhao et al. \cite{Zhao:scr16} were corrected for spin pumping but not for radiative effects.

\section{Conclusion}
We have developed a method to calculate the scattering matrix including SOC and non-collinearity and illustrated it here with calculations of the resistivity and Gilbert damping of permalloy. The very efficient implementation with tight-binding muffin tin orbits allows it to be applied to a wide range of materials and systems. In addition to zero-temperature calculations for ideal disordered alloys \cite{Starikov:prl10}, the method can be used to model temperature-induced disorder \cite{LiuY:prb11, LiuY:prb15}, for systems such as interfaces that are not periodic \cite{LiuY:prl14, WangL:prl16}, or for non-collinear systems \cite{YuanZ:prl12, YuanZ:prl14}. Our results for disordered alloys are in agreement with or have been confirmed by other established theoretical methods like CPA. Where comparison can be made, our results are in good agreement with experiment so they can be used to predict material parameters. 

We acknowledge many useful discussions with Rien Wesselink and Kriti Gupta. This work was financially supported by the ``Nederlandse Organisatie voor Wetenschappelijk Onderzoek.'' (NWO) through the research programme of ``Stichting voor Fundamenteel Onderzoek der Materie,'' (FOM) and through the use of supercomputer facilities of NWO ``Exacte Wetenschappen'' (Physical Sciences). It was also supported by EU Contract No. IST-033749 DynaMax, EU FP7 Contract No. NMP3-SL-2009-233513 MONAMI, ICT Grant No. 251759 MACALO, the Royal Netherlands Academy of Arts and Sciences (KNAW) and ``NanoNed,'' a nanotechnology programme of the Dutch Ministry of Economic Affairs.

\appendix

\section{Computational details}
\label{sec:cdappendix}
\subsection{Linearised Muffin-Tin Orbitals (LMTOs) and the Pauli-Schr{\"o}dinger Hamiltonian}
\label{sec:lmtosoc}

The practical implementation of the WFM method presented in this paper is based on LMTOs and ASA that are described in detail elsewhere \cite{Andersen:prl84, Andersen:85, Andersen:prb86}. Here we will focus only on those aspects which are important for the present version of the scattering formalism. An earlier, non-relativisitic version of the method \cite{Xia:prb06, Zwierzycki:pssb08} can be referred to for additional details regarding the use of muffin-tin orbitals in transport calculations.

In the ASA, muffin-tin spheres are expanded to fill the volume of the solid \cite{Andersen:prb75}. Considering only the $l=0$, spherically symmetric component of the potential inside the resulting AS located at ${\bf R}$, a solution  $\phi_{{\bf R} l}(E, r_{\bf R})$ of the radial Schr{\"{o}}dinger equation can be determined numerically for energy $E$ and angular momentum $l$ resulting in the partial wave \cite{Andersen:prb86}
\begin{equation}
\phi_{{\bf R} L}(E, {\bf r}_{\bf R}) \equiv \phi_{{\bf R} l}(E, r_{\bf R}) Y_{lm} (\hat{\bf r}_{\bf R})\:,
\end{equation}
with ${\bf r_R \equiv r-R}$, $L \equiv lm$ labels the angular momentum and $Y_{lm}$ is a spherical (or real, cubic) harmonic. $\hat{\bf r}_{\bf R}$ denotes a unit vector and $r_{\bf R} \equiv |{\bf r-R}|$. Later we can drop the explicit ${\bf R}$-dependence where it does not give rise to ambiguity. In terms of the logarithmic derivative $D_l$ of $\phi_{l}(E,r)$ at $r\equiv s$ (the AS radius)
\begin{equation}
  D_l(E, s) \equiv 
  \left.\frac{r \phi_l^{\prime}(E, r)}{\phi_l(E, r)}\right|_{r=s} \equiv 
  \frac{s\phi_l^{\prime}(E, s)}{\phi_l(E, s)}\;,
\end{equation}
we can define the energy-dependent potential function
\begin{equation}
  P_l^0(E)=2(2l+1) {\left(\frac{w}{s}\right)}^{2l+1}
  \frac{D_l(E)+l+1}{D_l(E)-l}\;,
  \label{eq:potfun0}  
\end{equation}
which in the ASA depends only on the potential and is independent of the crystal structure. When there is more than one atom type, $w$ is an average AS radius. In terms of the ``canonical'' structure constant matrix ${\bf S}^0_{{\bf R}^{\prime} L^{\prime}, {\bf R} L}$ \cite{Andersen:prb75, Andersen:prb86}, which depends only on the positions of the ions, we can define the Hermitian matrix at $E=E_{\nu}$ as
\begin{equation}
  {\bf h}^{0}(E_{\nu})=-{\left[\dot{{\bf P}}^{0}(E_{\nu})\right]}^{-1/2}\left[{\bf P}^{0}(E_{\nu})-{\bf S}^{0}\right]{\left[\dot{{\bf P}}^{0}(E_{\nu})\right]}^{-1/2} \;,
\end{equation}
where $P^0_{{\bf R} L}$ is an ($m$-independent) element of the diagonal matrix ${\bf P}^0$ and $\dot{{\bf P}}^{0}$ is the energy-derivative of ${\bf P}^0$. In this expression only ${\bf S}^{0}$ is a non-diagonal matrix. To first order in $E-E_{\nu}$, ${\bf h}^0(E_{\nu})$ is the Hamiltonian in the ASA \cite{ Andersen:prl84, Andersen:85, Andersen:prb86}. The problem presented by the long range \cite{Andersen:prb75} of the structure constant matrix ${\bf S}^0$ is resolved by introducing a generalised representation characterised by a set of $l$-dependent screening parameters $\alpha_l$ and defining so-called ``screened'' structure constants ${\bf S}^{\alpha}$ and potential functions ${\bf P}^{\alpha}$ defined by
\begin{align}
    {\bf P}^{\alpha}&={\bf P}^{0}{\left[1-{\bm\alpha} {\bf P}^{0}\right]}^{-1}\;,\label{eq:potentialfunc}\\
    {\bf S}^{\alpha}&={\bf S}^{0}{\left[1-{\bm\alpha} {\bf S}^{0}\right]}^{-1}\;.    
\end{align}
where $\bm\alpha$ is a diagonal matrix with $m$-independent elements $\alpha_l$. For a suitable choice of screening parameters, the range of ${\bf S}^{\alpha}$ is essentially limited to the first- and second-nearest neighbours for close-packed structures \cite{Andersen:prl84, Andersen:85, Andersen:prb86}. In the screened representation, the two-center tight-binding matrix becomes 
\begin{equation}  
{\bf h}^{\alpha}(E_{\nu})=-{\left[\dot{{\bf P}}^{\alpha}(E_{\nu})\right]}^{-1/2}
                   \big[{\bf P}^{\alpha}(E_{\nu})-{\bf S}^{\alpha}\big]
              {\left[\dot{{\bf P}}^{\alpha}(E_{\nu})\right]}^{-1/2} .
\end{equation}

The energy-independent, linearised (at $E_{\nu}$) muffin-tin orbitals for the AS located at ${\bf R}$ are defined \cite{Andersen:prb86} as
\begin{equation}
    \big|\chi^{\alpha}_{RL}(E_{\nu})\big\rangle=
    \big|\phi_{RL}(E_{\nu})\big\rangle+
    \big|\dot{\phi}^{\alpha}_{R'L'}(E_{\nu})\big\rangle 
                    h^{\alpha}_{R'L',RL}(E_{\nu}) \;,
\label{eq:LMTO}
\end{equation}
where
\begin{equation}
\big|\dot{\phi}^{\alpha}_{R'L'}(E_{\nu})\big\rangle=\frac{1}{N^{\alpha}_l(E_{\nu})}\left.\frac{\partial\left[\big|\phi_{R'L'}(E)\big\rangle N^{\alpha}_l(E)\right]}{\partial E}\right|_{E=E_{\nu}},
\label{eq:mtophidot}
\end{equation}
with the normalization function
\begin{equation}
N^{\alpha}_l(E_{\nu})={\left[(s/2) \dot{P}^{\alpha}_l(E_{\nu})\right]}^{1/2}\;.
\label{eq:mtonorm}
\end{equation}
For simplicity, we will later assume that the orbitals are constructed at $E_{\nu}$ and omit any explicit energy dependence. Now we can construct the energy independent Hamiltonian matrix correct to second order in $E-E_{\nu}$
\begin{equation}
  \langle  {\bf \chi}^{\alpha}|{\bf H}-E_{\nu}\mathbf I |{\bf \chi}^{\alpha}\rangle = {\bf h}^{\alpha} + {\bf h}^{\alpha} {\bf o}^{\alpha} {\bf h}^{\alpha} \;,  
\label{eq:lmtoham}
\end{equation}
where ${\bf o}^{\alpha}=\langle {\bm \phi} | \dot{\bm \phi}^{\alpha}\rangle=\dot{\bf N}^{\alpha}/{\bf N}^{\alpha}$ is the so-called overlap matrix. Equation \eqref{eq:lmtoham} shows that the hopping range of the LMTO Hamiltonian is double the hopping range of the screened structure constant matrix, defined by the three-center integral ${\bf h}^{\alpha} {\bf o}^{\alpha} {\bf h}^{\alpha} $. In a transport calculation dominated by what happens at the Fermi energy $E_F$ we can choose $E_{\nu}=E_F$ and the second (three-centre) term in Eq. (\ref{eq:lmtoham}) can be omitted.

We include the spin-orbit interaction in a perturbative way by adding a Pauli term to the Hamiltonian 
\begin{equation}
 \mathbf H_{\rm so}=\frac{1}{c^2\: r} \frac{dV(r)}{dr} \: \hat{{\bf L}} \cdot \hat{{\bf S}}\;. 
\label{eq:hso}
\end{equation}
In the LMTO basis set, the matrix elements of $H_{\rm so}$ are given by
\begin{align*}
\langle{\bf \chi}^{\alpha} | \mathbf H_{\rm so} |  {\bf \chi}^{\alpha} \rangle
  = {\bm \gamma}_1 
  + {\bm \gamma}_2  {\bf h}^{\alpha}+ {\bf h}^{\alpha}{\bm \gamma}^+_2 
  + {\bf h}^{\alpha} {\bm \gamma}_3  {\bf h}^{\alpha}  ,
\end{align*}
where $\bm\gamma_1$, $\bm\gamma_2$ and $\bm\gamma_3$ are spin-orbit parameters for one-, two-, and three-center terms
\begin{subequations}
\begin{alignat}{3}
{\bm \gamma}_1 &= \bigg\langle{\bm \phi}
  &&{ \bigg| \frac{1}{c^2\: r} \frac{dV(r)}{dr} \: \hat{\bf L} \cdot \hat{\bf S} \bigg|}
                                                                {\bm \phi}\bigg\rangle 
  &&= {\bf K} \otimes {\bm \xi}  \\
{\bm \gamma}_2 &= \bigg\langle{\bm \phi}
  &&{ \bigg| \frac{1}{c^2\: r} \frac{dV(r)}{dr} \: \hat{\bf L} \cdot \hat{\bf S} \bigg|}
                                                   \dot{\bm \phi}^{\alpha}\bigg\rangle 
  &&= {\bf K} \otimes \dot{\bm \xi}^{\alpha}  \\ 
{\bm \gamma}_3 &= \bigg\langle\dot{\bm \phi}^{\alpha}
  &&{ \bigg| \frac{1}{c^2\: r} \frac{dV(r)}{dr} \: \hat{\bf L} \cdot \hat{\bf S} \bigg|} 
                                                  \dot{\bm \phi}^{\alpha} \bigg\rangle 
  &&= {\bf K} \otimes \ddot{\bm \xi}^{\alpha}       
\end{alignat}
\label{eq:SOCparameters}     
\end{subequations}
and we introduce a matrix of coefficients 
\begin{equation}
  K_{lm\sigma,l'm'\sigma'}=\langle lm\sigma | \hat{\bf L} \cdot \hat{\bf S}|l'm'\sigma' \rangle ,
\label{eq:lscoeffs}
\end{equation}
and set of SOC potential parameters
\begin{subequations}
\begin{align}
     {\xi}_{l\sigma\sigma'} 
  &=\left\langle \phi_{l\sigma}(r)
    \left|\frac{1}{c^2\: r} \frac{dV^{\sigma\sigma'}(r)}{dr} \right|\phi_{l\sigma'}(r)\right\rangle \\
 \dot{\xi}_{l\sigma\sigma'}^{\alpha} 
  &=\left\langle \phi_{l\sigma}(r)
    \left|\frac{1}{c^2\: r} \frac{dV^{\sigma\sigma'}(r)}{dr} \right|\dot{\phi}^{\alpha}_{l\sigma'}(r)\right\rangle \\
\ddot{\xi}_{l\sigma\sigma'}^{\alpha} 
  &=\left\langle \dot{\phi}^{\alpha}_{l\sigma}(r)
    \left|\frac{1}{c^2\: r} \frac{dV^{\sigma\sigma'}(r)}{dr} \right|\dot{\phi}^{\alpha}_{l\sigma'}(r)\right\rangle \;.   
\end{align}
\end{subequations}
The expressions for $\dot{\xi}$ and $\ddot{\xi}$ can be slightly reworked by taking  \eqref{eq:mtophidot} into account
\begin{subequations}
\begin{align}
  \dot{\xi}_{l\sigma\sigma'}^{\alpha} &=  \dot{\xi}_{l\sigma\sigma'} + 
      {\xi}_{l\sigma\sigma'} o^{\alpha}_{l\sigma'}   \\
 \ddot{\xi}_{l\sigma\sigma'}^{\alpha} &= \ddot{\xi}_{l\sigma\sigma'} +
  \dot{\xi}_{l\sigma\sigma'} \left(o^{\alpha}_{l\sigma}+ o^{\alpha}_{l\sigma'}\right)+
      {\xi}_{l\sigma\sigma'} o^{\alpha}_{l\sigma} o^{\alpha}_{l\sigma'} ,  
\end{align}
\end{subequations}
where 
\begin{subequations}
\begin{align}
{\dot{\xi}}_{l\sigma\sigma'}=& \left\langle\phi_{l\sigma}(r)\left|\frac{1}{c^2\: r} \frac{dV^{\sigma\sigma'}(r)}{dr} \right|{\dot{\phi}}_{l\sigma'}(r)\right\rangle \\
{\ddot{\xi}}_{l\sigma\sigma'} =& \left\langle{\dot{\phi}}_{l\sigma}(r) \left|\frac{1}{c^2\: r} \frac{dV^{\sigma\sigma'}(r)}{dr} \right|{\dot{\phi}}_{l'\sigma'}(r)\right\rangle .
\end{align}
\end{subequations}
The parameters $\xi$, $\dot{\xi}$ and $\ddot{\xi}$ can be obtained by radial integration over the AS. Off-diagonal (in spin-space) elements of $V^{\sigma\sigma'}(r)$ are assumed to be the average of potentials for different spins \cite{Daalderop:prb90a}.

\begin{figure}[t]
  \centering
  \includegraphics[width=0.45\textwidth]{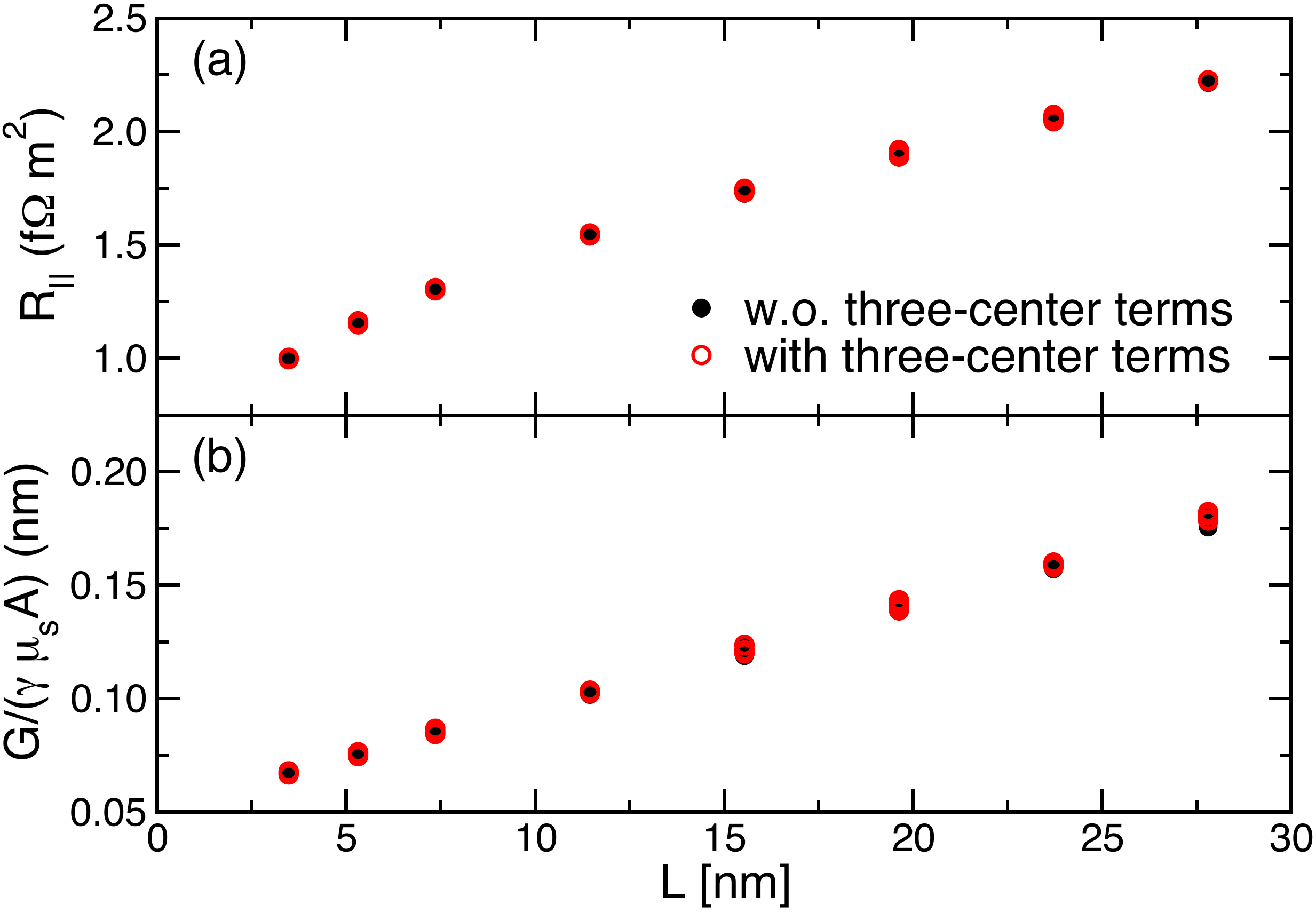}  
\caption{Effect of the three center terms in ${\bf H}_{\rm so}$ on (a) the total resistance and (b) the total damping of $\rm Ni_{80}Fe_{20}$ at $T=0$. Red open circles include three-center terms, black filled circles omit them. 
}
\label{fig:3center}
\end{figure}

Thus, the complete Pauli-Schr{\"{o}}dinger Hamiltonian will be 
\begin{equation}
  {\bf H}_{\rm so}= \big[E_{\nu} \mathbf I+{\bm \gamma}_1 \big]
 + \big[{\bf h}^{\alpha}+{\bm \gamma}_2 {\bf h}^{\alpha}+ {\bf h}^{\alpha}{\bm \gamma}^+_2 \big] 
 + \big[ {\bf h}^{\alpha} ({\bf o}^{\alpha}+{\bm \gamma}_3) {\bf h}^{\alpha}\big],
  \label{eq:socham}
\end{equation}
where we have grouped one-, two- and three-centre terms. Even when $E=E_{\nu}$, omitting the three-centre term in (\ref{eq:socham}) is formally less readily justified than when SOC is neglected because it introduces longer-range hopping in the Hamiltonian matrix. The practical impact on the resistivity and damping of neglecting the three-centre terms is examined in Fig.~\ref{fig:3center} where no significant effect can be seen while the computational cost is reduced by some 70\%. 

\subsection{Velocities}
\label{sec:veloc}
In this section we derive the expression \eqref{ando:veloc} for the group velocities of eigenmodes in the ideal wire for the generalised WFM framework.
The derivations are similar to previous work \cite{Khomyakov:prb04, Khomyakov:prb05}. The vectors ${\bf u}_m$ of \eqref{eq:uvec} are solutions of the polynomial equation of order $2N$
\begin{equation}
\lambda ^N {\bf H} {\bf u}_m  + \sum_{n=1}^N \left( \lambda_m^{N+n}  {\bf B}_n {\bf u}_m + \lambda_m^{N-n}  {\bf B}^{\dagger}_n {\bf u}_m\right) =0\;.
\label{eq:apv1}
\end{equation}
Left-multiplying by ${\bf u}_m^{\dagger}$ and differentiating with respect to energy leads to
\begin{align}
\frac{d}{dE}\!\!\! \left[\!
\lambda_m^N {\bf u}_m^{\dagger}{\bf H} {\bf u}_m \! + \!\!\sum_{n=1}^N \!\left( \lambda_m^{N+n} {\bf u}_m^{\dagger} {\bf B}_n {\bf u}_m\! + \!\lambda_m^{N-n} {\bf u}_m^{\dagger} {\bf B}^{\dagger}_n {\bf u}_m\right) \!\right]\!  
\nonumber\\
=\lambda_m^N +
2 \imath \lambda_m^{N-1} \frac{d\lambda_m}{dE} 
\sum_{n=1}^N n \:\mathrm{ Im}\left( \lambda_m^{n} {\bf u}_m^{\dagger} {\bf B}_n {\bf u}_m \right)
 = 0 \;,
\end{align}
which yield the following expression for $dE/d\lambda_m$: 
\begin{equation}
\frac{dE}{d\lambda_m}=-\frac{2 \imath}{\lambda_m} \sum_{n=1}^N n \:\mathrm{ Im}\left( \lambda_m^{n} {\bf u}_m^{\dagger} {\bf B}_n {\bf u}_m \right)\;.
\label{eq:apdedl}
\end{equation}
For propagating states, $\lambda_m=e^{ik_m a}$, $k_m$ is real and $a$ is the thickness of the periodic lead layer so
\begin{equation}
  \frac{dk_m}{dE}=\frac{1}{\imath a \lambda_m}\frac{d\lambda_m}{dE}\;.
  \label{eq:dkde}
\end{equation}
The standard definition of group velocity is $\upsilon_m=\frac{1}{\hbar} \frac{dE}{dk}$, therefore, substituting Eq. (\ref{eq:apdedl}) and (\ref{eq:dkde}), we obtain
\begin{equation}
\upsilon_m=\frac{\imath a \lambda_m}{\hbar} \frac{dE}{d\lambda_m}= \frac{2a}{\hbar} \sum_{n=1}^N n \:\mathrm{ Im}\left( \lambda_m^{n} {\bf u}_m^{\dagger} {\bf B}_n {\bf u}_m \right).
\label{eq:apvel}
\end{equation}

\subsection{Spin-projections of the scattering matrix}
\label{sec:decompose}
For convenience of interpretation it is useful to decompose the transmission and reflection matrices into spin-projected ones. Although spin is not a valid quantum number when SOC is included, we can still characterise states in the leads as states which have a distinctive spin projection onto the $\sigma_{z}$ axis ($\sigma_{z}=\pm \frac{1}{2}$). For leads consisting of light non-magnetic metals, this is a reasonable approximation and can be achieved in practice as follows: for every pair of spin-degenerate lead eigenmodes ${\bf u}_1,{\bf u}_2$ we can construct a new pair of orthogonal eigenmodes ${\bf u}'_1,{\bf u}'_2$ by taking a linear superposition of the original modes
\begin{equation}
  \left[
  \begin{array}{c}
  {\bf u}'_1 \\
  {\bf u}'_2 
  \end{array}
  \right]
  =
  \left[
  \begin{array}{cc}
  a_{11} & a_{12}\\
  a_{21} & a_{22}
  \end{array}
  \right]
  \times
  \left[
  \begin{array}{c}
  {\bf u}_1\\
  {\bf u}_2
  \end{array}
  \right]    
\end{equation}
and choosing the coefficients $a_{ij}$ to maximize the $\sigma_z=+\frac{1}{2}$ component of ${\bf u}'_1$  and the $\sigma_z=-\frac{1}{2}$ component of ${\bf u}'_2$. We denote these new states as ${\bf u}^{\sigma+}$ and ${\bf u}^{\sigma-}$. This basis set transformation allows us to to operate with reasonably well defined spin-projected scattering matrices. For example, the matrix $r^{\sigma\sigma^{\prime}}_{\mu\nu}$ describes reflection from the $\nu\sigma'$ states into the $\mu\sigma$ states in the same lead.

\subsection{Reduction of the Gilbert damping tensor to a scalar}
\label{sec:isotropic}
\begin{figure}[h]
\begin{center}
\includegraphics[width=0.9\columnwidth]{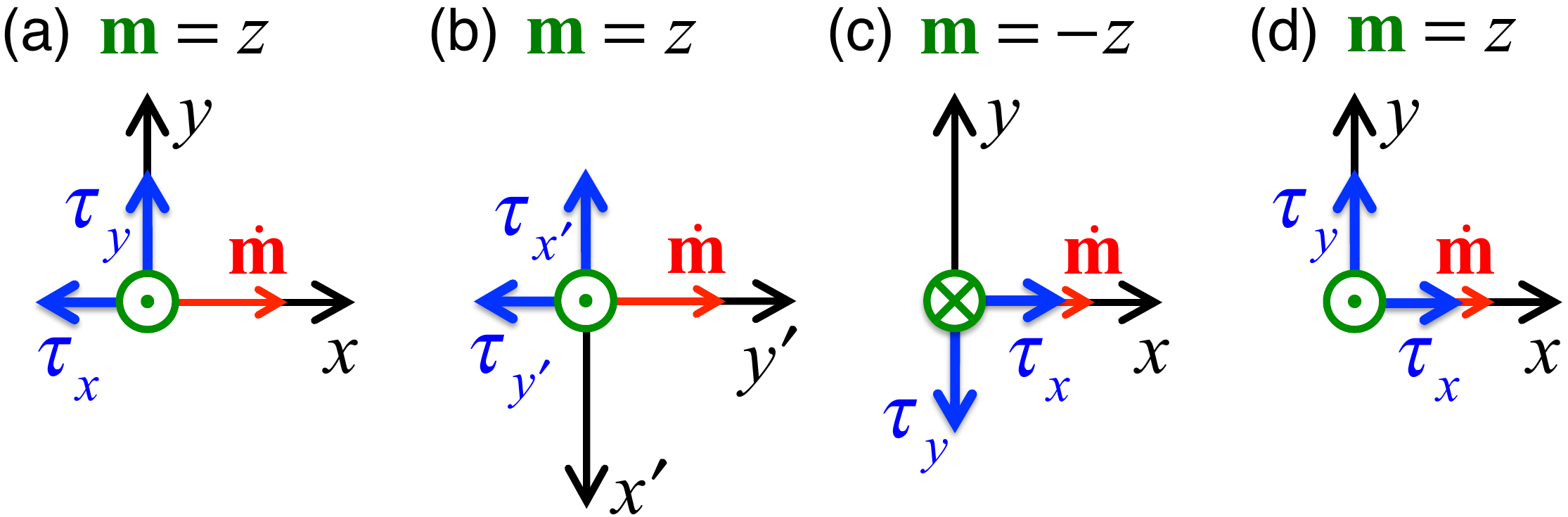}
\caption{Geometry of damping torque exerted on magnetization. See text for detailed analysis.}
\label{fig1}
\end{center}
\end{figure}

For a crystal with cubic symmetry and a collinear magnetisation, the damping torque can in general be written as ${\bm \tau} = {\bf m} \times ({\bm \alpha} \cdot {\bf \dot  m} )$. If the magnetisation $\mathbf m$ is taken to be along the $z$-axis, then to leading order in the transverse components of the magnetisation ($m_x, m_y \ll m_z \sim 1$) the damping torque can be written in cartesian coordinates as
\begin{subequations}
\begin{alignat}{2}	
\tau_x&=-&m_z\left(\alpha_{yx}\dot m_x+\alpha_{yy}\dot m_y\right),\\
\tau_y&= &m_z\left(\alpha_{xx}\dot m_x+\alpha_{xy}\dot m_y\right).
\end{alignat}	
\end{subequations}
Without loss of generality, we choose the momentary direction of magnetisation precession to be along the $x$-axis i.e., $\dot{\mathbf m}=\dot m\hat x$ and $\vert m_z\vert =1$, as sketched in Fig.~\ref{fig1}(a). Then $\tau_x=-\alpha_{yx}\dot m$ and $\tau_y=\alpha_{xx}\dot m$. Keeping the system otherwise unchanged, we rotate the coordinate axes through 90$^\circ$ clockwise about the $z$-axis  as shown in Fig.~\ref{fig1}(b). In this case we can write the damping torque $\tau_{x'}=-\alpha_{yy}\dot m$ and $\tau_{y'}=\alpha_{xy}\dot m$. Comparing the components in Fig.~\ref{fig1}(a) and (b), we find $\alpha_{xx}=\alpha_{yy}$ and $\alpha_{xy}=-\alpha_{yx}$.

If we reverse the magnetization in Fig.~\ref{fig1}(a) but keep $\dot m$ unchanged, the damping torque should be reversed as plotted in Fig.~\ref{fig1}(c). If we then rotate the system 180$^\circ$ about the $x$-axis, it reproduces the configuration of Fig.~\ref{fig1}(a) except that the component $\tau_x$ is inverted. As a consequence, $\tau_x=-\alpha_{yx}\dot m$ must be zero indicating that the off-diagonal elements $\alpha_{yx}=-\alpha_{xy}=0$. In the same manner, it can be proved that in a collinear magnetic system with cubic symmetry, the Gilbert damping tensor reduces to a scalar, $\bm{\alpha}=\alpha\mathbf 1$, where $\mathbf 1$ is the $3\times3$ unit matrix.

\section{Limitations}
\label{sec:limitations}

A number of factors limit the application of the present method. First and foremost, memory considerations limit the size of system that can be addressed. A calculation for a single configuration of the longest scattering region shown in Fig.~\ref{fig:resistance} and containing about $15,000 =5 \times 5 \times 600$ atoms requires about one hour on a supercomputer node with 32 cores and 256 GB memory, where the calculation parallelized perfectly over the two dimensional 32$\times$32 $k$-point summation. The maximum length of $\sim$105~nm places an upper limit on e.g. the spin-flip diffusion length that could be studied. For larger systems, the calculations need to be performed with a multithreading sparse matrix solver or simply with extended memory. For metals, the computing time scales linearly with the length $L$ of the scattering region and quadratically with the size of the lateral supercell. For a lateral supercell containing $M$ atoms a calculation for a single $k$-point scales as $M^3$. The BZ size scales as $1/M$ so for constant sampling density of reciprocal space, the scaling goes as $M^2L$. The upper limit of lateral supercell size with SOC is in practice about 10$\times$10 and 30~nm long; this scattering region also contains about 15,000 atoms. 

A second important limitation is the ASA. This conventional description of the potential in combination with the MTO scheme is usually very good for close-packed systems; open systems can frequently be reasonably well modelled by filling space with artificial ``empty atomic spheres'' \cite{Glotzel:ssc80, Andersen:prb86}, i.e. without nuclei inside. For lower symmetry structures, where the spherically symmetric potentials around the nuclei are not sufficient to characterize the real Kohn-Sham potential, the ASA breaks down. In these cases, the reliability and accuracy of transport calculations as currently implemented with MTO and ASA are limited by the ASA description of the Kohn-Sham potentials. Andersen has suggested ways of circumventing this limitation without sacrificing the efficiency of the ASA \cite{Zwierzycki:appa09}.

The ultimate limitation is the DFT itself, or rather the approximation to the exchange-correlation potential, the functional derivative of the exchange-correlation (XC) energy, that has to be chosen. Some of the uncertainties that result from difference choices are discussed briefly next.

\subsection*{Additional numerical tests}
\label{sec:extratests}

Although the formalism described in this paper is parameter free, the practical implementation requires approximating the XC energy of DFT \cite{Hohenberg:pr64} as in the local density approximation \cite{Kohn:pr65} where electron gas data have been parameterized by various workers \cite{vonBarth:jpc72, Perdew:prb81, Vosko:cjp80}. The numerical values of the resistivity and damping parameter will depend on the parameterization chosen.  Our implementation with TB-MTOs requires truncating the orbital angular momentum expansion at some maximum value of the orbital angular momentum and this will also influence the numerical results. For all calculations presented in this paper, the von Barth-Hedin (vBH) XC potential and an $spd$ basis set were used. To demonstrate the uncertainty resulting from the somewhat arbitrary choice of XC potential and basis sets,  we show the results of calculations for $\rm Cu|Py|Cu$ with different choices of potentials and $spd$ or $spdf$ basis set in Table~\ref{tab:conv}.

\begin{table}[t]
  \caption{Dependence of the atomic magnetic moment $\mu_s$, non-relativistic resistivities $\rho_{\rm maj}$ and $\rho_{\rm min}$, the relativistic resistivity $\rho_{\parallel}$, and the Gilbert damping parameter for $\rm Py$, on the basis set and choice of exchange-correlation potential: von Barth-Hedin (vBH) \cite{vonBarth:jpc72}, Perdew-Zunger (PZ) \cite{Perdew:prb81} and Vosko-Wilk-Nusair (VWN)\cite{Vosko:cjp80}. Calculations are performed with a $5\times5$ supercell and a $k$-point sampling grid of $32\times32$ (equivalent to $160\times160$ for a $1\times1$ primitive interface cell).
Resistivities are given in $\mu\Omega \,$cm and magnetic moment is in Bohr magneton $\mu_B$ per atom.
  }. 
  \begin{center}
    \begin{tabular*}{0.49\textwidth}{@{\extracolsep{\fill}}l c c c c c}
      \hline \hline
      XC/Basis & $\mu_s$ & $\rho_{\rm maj} $& $\rho_{\rm min} $& $\rho_{\parallel}$ &$\alpha$ (10$^{-3}$) \\    
      \hline
      vBH/$spd$  & $1.025$ & $0.57\pm 0.01$ & $109\pm 1$ &  $2.7\pm 0.1$ & $4.6\pm0.2$ \\
      vBH/$spdf$ & $1.001$ & $0.67\pm 0.01$ & $101\pm 1$ &  $2.6\pm 0.1$ & $4.3\pm0.2$ \\    
      PZ/$spd$    & $1.010$ & $0.92\pm 0.01$ & $108\pm 1$ &  $3.1\pm 0.1$ & $4.7\pm0.2$ \\    
      VWN/$spd$ & $1.022$ & $0.60\pm 0.01$ & $107\pm 1$ &  $2.8\pm 0.1$ & $4.5\pm0.2$ \\    
      \hline \hline
    \end{tabular*}
  \end{center}
\label{tab:conv}  
\end{table}

The quantity most dependent on these choices is the permalloy majority spin resistivity in the absence of SOC. The very weak scattering of majority spins makes this very sensitive to small details of the electronic structure which in turn depend strongly on the exchange splitting. Once SOC is included, the mean-free path is reduced and the sensitivity of the resistivity and, especially, of the Gilbert-damping parameter to these ``technical'' details becomes acceptable.

Various forms of the XC potentials have been implemented in computer programs and examined for different physical and chemical quantities. \cite{Martin:04} For the transport properties of magnetic metals and alloys that we study in this paper, there is no clear evidence to show that one XC potential is better than the others. The test for the basis set using the vBH XC potential also indicates that the minimal $spd$ basis is good enough for convergence, as initially proposed by Lambrecht and Andersen \cite{Lambrecht:prb86}. The slight difference in the calculated resistivity and Gilbert damping can be regarded as the ``uncertainty'' arising from an arbitrary choice of XC potential.

\bibliography{pjk,notes}

\begin{thebibliography}{149}
\bibitem [{foo()}]{footnote1}%
  \BibitemOpen
  \href@noop {} {}\bibinfo {note} {The finite intercept at $L=0$ in
  Fig.~\ref{fig:damppy}, $\widetilde G_i \equiv \widetilde G(L=0)$ corresponds
  to an interface damping enhancement that was observed for thin ferromagnetic
  films \cite{Heinrich:prl87, Platow:prb98, Urban:prl01} and NM$|$Py$|$NM
  layers \cite{Mizukami:jmmm01, *Mizukami:jjap01, Mizukami:jmmm02,
  Ingvarsson:prb02}. It was qualitatively explained in terms of how a
  precessing magnetization ``pumps'' the transverse component of magnetization
  through the interface \cite{Tserkovnyak:prl02a, *Tserkovnyak:prb02b}. The
  material parameter that describes this ``spin pumping'' is the ``mixing
  conductance'' introduced to describe spin dependent transport between
  non-collinearly aligned elements of a magnetic circuit \cite{Brataas:prl00,
  *Brataas:epjb01}. To describe the interface damping enhancement
  quantitatively, diffusive scattering and spin-flip scattering in the NM layer
  have to be described as well as interface spin-flip scattering and
  modification of the mixing conductance by SOC \cite{LiuY:prl14}
  .}\BibitemShut {Stop}%
\end{thebibliography}

\end{document}